\documentclass[manuscript,authorversion,review=false,timestamp=false]{acmart}

\usepackage{caption}
\usepackage{subcaption}
\usepackage{tabularx}
\usepackage{colortbl}
\usepackage{xcolor}
\usepackage{placeins}
\usepackage{longtable}
\usepackage{balance}
\usepackage{booktabs}

\usepackage{float}
\floatstyle{plaintop}
\restylefloat{table}

\AtBeginDocument{%
  \providecommand\BibTeX{{%
    \normalfont B\kern-0.5em{\scshape i\kern-0.25em b}\kern-0.8em\TeX}}}

\begin{document}

\title[VSD Guidelines]{Guidelines for Integrating Value Sensitive Design in Responsible AI Toolkits}

\author{Malak Sadek}
\email{m.sadek21@imperial.ac.uk}
\orcid{0000-0001-8284-890X}
\affiliation{%
  \institution{Imperial College London}
  \streetaddress{Exhibition Road}
  \city{London}
  \country{UK}
  \postcode{SW7 2BX}
}

\author{Marios Constantinides}
\email{marios.constantinides@nokia-bell-labs.com}
\orcid{0000-0003-1454-0641}
\affiliation{%
  \institution{Nokia Bell Labs}
  \city{Cambridge}
  \country{UK}
  \postcode{CB3 0FA}
}

\author{Daniele Quercia}
\email{daniele.quercia@nokia-bell-labs.com}
\orcid{0000-0001-9461-5804}
\affiliation{%
  \institution{Nokia Bell Labs}
  \city{Cambridge}
  \country{UK}
  \postcode{CB3 0FA}
}

\author{Céline Mougenot}
\email{c.mougenot@imperial.ac.uk}
\orcid{0000-0002-3849-163X}
\affiliation{%
  \institution{Imperial College London}
  \streetaddress{Exhibition Road}
  \city{London}
  \country{UK}
  \postcode{SW7 2BX}
}

\renewcommand{\shortauthors}{Sadek et al.}

\begin{abstract}

Value Sensitive Design (VSD) is a framework for integrating human values throughout the technology design process. In parallel, Responsible AI (RAI) advocates for the development of systems aligning with ethical values, such as fairness and transparency. In this study, we posit that a VSD approach is not only compatible, but also advantageous to the development of RAI toolkits. To empirically assess this hypothesis, we conducted four workshops involving 17 early-career AI researchers. Our aim was to establish links between VSD and RAI values while examining how existing toolkits incorporate VSD principles in their design. Our findings show that collaborative and educational design features within these toolkits, including illustrative examples and open-ended cues, facilitate an understanding of human and ethical values, and empower researchers to incorporate values into AI systems. Drawing on these insights, we formulated six design guidelines for integrating VSD values into the development of RAI toolkits.

\end{abstract}




\maketitle

\section{Introduction}
\label{sec:introduction}

The increase of risks associated with Artificial Intelligence (AI) systems~\citep{buolamwini2018, fox2023} have led to a surge in the development of toolkits aimed at facilitating the practical design of Responsible AI (RAI) \citep{wong2022}. RAI advocates for the responsible design, development, and use of AI systems, aligning with values like fairness and transparency ~\cite{tahaei2023human}. A framework with similar objectives is Value Sensitive Design (VSD), recognised for creating more human-centred AI systems \citep{umbrello2021}. 
This study hypothesises that VSD can effectively guide the creation of RAI toolkits, given its consideration of human values in the technology design process. However, the extent of alignment between VSD and RAI values and how existing RAI toolkits incorporate VSD remain unclear. Two critical aspects for supporting VSD involve: \emph{a)} enabling stakeholder inclusion and collaboration in order to elicit and understand their values \citep{manders2009, sadek2023AE}, and \emph{b)} facilitating the education of toolkit users in order to promote self-reflexivity and responsible decision-making \citep{morley2021, umbrello2021, garst2022}.

This paper focuses on the the practical application of VSD in RAI toolkits, in two Research Questions (RQs):

\begin{itemize}
    \item \textbf{RQ\textsubscript{1}:} How closely do Value Sensitive Design (VSD) values align with Responsible AI (RAI) values?
    \item \textbf{RQ\textsubscript{2}:} How do existing RAI toolkits incorporate VSD, and support collaboration and learning?
\end{itemize}

To address these questions, we conducted an empirical investigation through workshops with 17 participants (AI researchers) (\S\ref{sec:methodology}). The workshops focused on the expression of VSD values within RAI toolkits, and the impact of toolkit design features on participants' perceptions of stakeholder collaboration and learning. 

The contributions of this study are threefold: (i) \textbf{mapping VSD values} onto commonly used RAI values integrated into the toolkits, revealing a high degree of alignment between the two sets of values evidenced by consensus among workshop participants (\S\ref{sec: RQ1}), (ii) \textbf{identifying key links} between design features in RAI toolkits and their impact on promoting VSD. This included: navigation methods supporting iteration, open-ended cuing supporting collaboration versus solo work, examples and case studies providing learning opportunities, and value incorporation reducing cognitive load (\S\ref{sec:RQ2}), and (iii) \textbf{formulating six practical design recommendations} for enhancing value sensitivity in RAI toolkits (\S\ref{sec:discussion}). These recommendations, focusing on concrete design features such as supporting actionability and shared knowledge, complement recent broader suggestions for the focuses and approaches of RAI toolkits \citep{elsayed2023}.
\section{Background and Related Work}
\label{sec:background}

\subsection{Value Sensitive Design and Alternative Frameworks}

\begin{figure*}[t!]
  \centering
  \includegraphics[width=\linewidth]{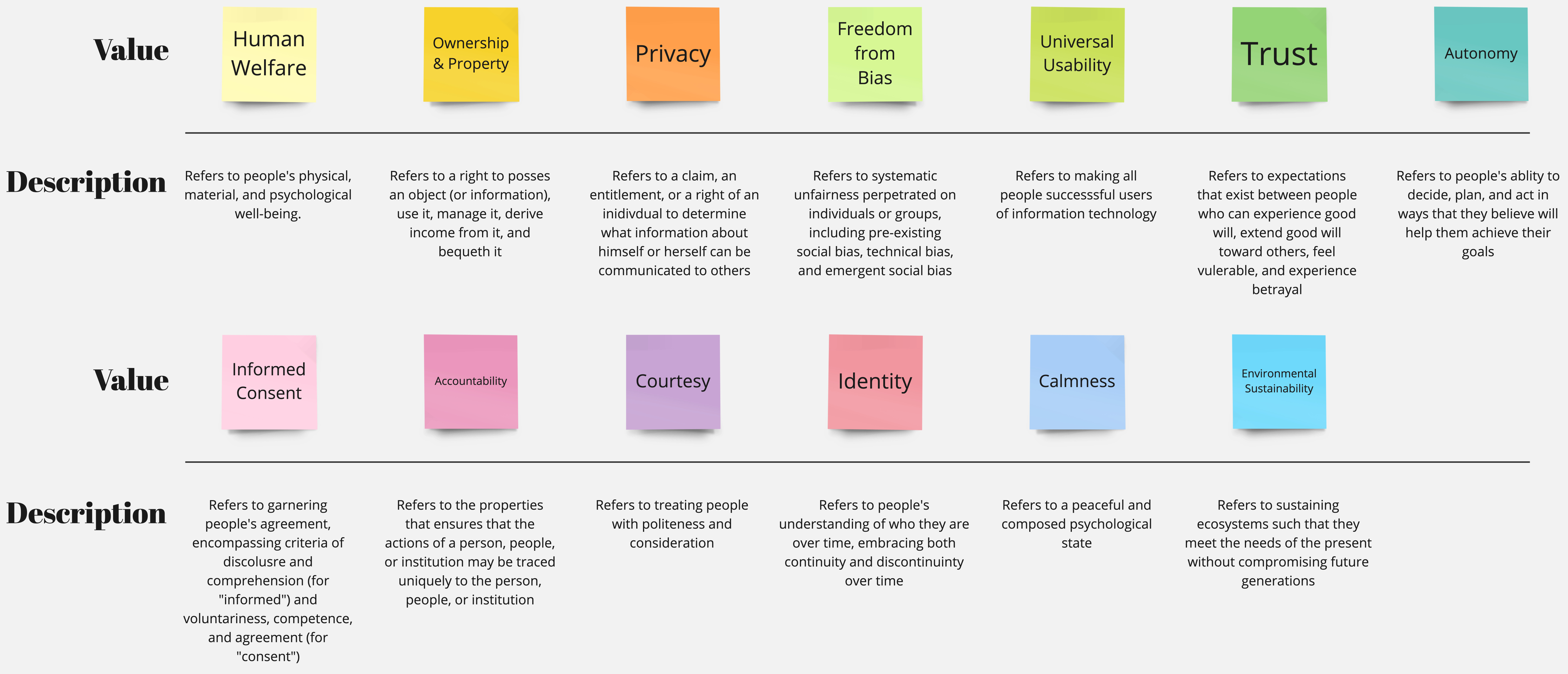}
  \caption{A list of the values stated by VSD as being ``often implicated in system design" \citep{friedman2002} and their descriptions in a Miro board.}
  \label{fig:vsd-values}
  \Description{A screenshot from a Miro board showing 13 coloured sticky notes with different values written on them and a text field showing a description of each values below its respective sticky note.}
\end{figure*}

VSD is a theoretical design framework which advocates for the elicitation and inclusion of stakeholders' values in technology~\citep{friedman2002, borning2012}. The framework outlines three types of investigations that allow this to happen: conceptual investigations focusing on identifying relevant stakeholders and understanding their context, empirical investigations aimed at understanding stakeholders' needs and values, and technical investigations to reflect on how the technology being created can enable or violate these values \citep{friedman2002, borning2005}. The framework of VSD is supportive of examining the role of values in emerging technologies \citep{berger2023} and in practice \citep{fuchsberger2012}, and embedding values in design collaborations \citep{yoo2013}. While there is strong support for eliciting contextual values directly from stakeholders for a given project \citep{manders2009}, VSD also offers a list of ``human values often implicated in system design" \citep{friedman2002}~[p.17] to consider. Figure \ref{fig:vsd-values} shows these values and their definitions exactly as stated by \citep{friedman2002} in a Miro board.

While several other ethical and value-based frameworks exist, such as Utilitarian or Egalitarian Ethics \citep{cherubini2017}, approaches where many people's values are aggregated, consolidated, or alternated have been recommended by experts in the context of AI systems \citep{cherubini2017, gabriel2020}. When coupled with (i) the consideration of technology generally \citep{poel2020}, and algorithms specifically, as ``value-laden artefacts" \citep{martin2019}, (ii) the tendency to focus solely on economic values in AI systems and the need to consider broader human values \citep{umbrelo2022a}, and (iii) the ability of VSD to promote self-reflexivity in AI practitioners \citep{umbrello2021} and practically bootstrap onto existing design processes for AI systems \citep{umbrello2021, umbrello2022b}, it becomes clear that VSD is an especially suitable framework to consider during AI system development. While design processes \cite{zhu2018, rahimi2019}, guideline \citep{boyd2022} and methodologies \cite{vera2019} for value-sensitive AI are emerging, the explicit focus on VSD when designing RAI toolkits remains limited.

\subsection{Responsible AI Toolkits}

The space of theoretical interventions for responsible AI, such as guidelines and recommendations, while crucial, is quickly becoming overwhelmingly saturated \citep{jobin2019}. Recent criticisms have expressed concern at the growing gap between theoretical interventions and the practical implementations of AI systems \citep{martinho2021, chen2022, jobin2019, digitalcatapult2020, custis2021}. Theoretical frameworks are being viewed as too abstract \citep{harbers2022}, difficult to practically interpret \citep{verbeek2021}, ineffective at resolving conflicts \citep{palmer2023}, offering little guidance \citep{whittlestone2019c}, immeasurable in terms of their impact \citep{hagendorff2020}, hindering accountability \citep{metcalf2021, moss2021}, and unimpactful on practitioners \citep{mcnamara2018}. 

As a result, there has been a shift towards practical tools and processes to guide the implementation of AI systems. These come in several forms such as software \cite{lee2019} and design methods \cite{vera2019}, activities \citep{delgado2022}, and toolkits \cite{mora2017}. The aim of practical interventions is to translate theoretical concepts and frameworks into a tangible, digestible form that practitioners can utilise within their workflows. Despite the basis of most of these practical interventions, and especially toolkits, on theoretical frameworks, very little work has been done to assess the extent to which they effectively operationalise the core concepts of those frameworks.

\subsubsection{Forms and Mediums}

RAI toolkits originate from both scholarly and industry-based sources. MIT's AI Blindspot cards \citep{MIT} and the Digital Impact toolkit by Stanford Digital Civil Society Lab \citep{stanford} are two examples of academic contributions, while Microsoft's Judgement Call cards \citep{microsoft} and Nokia AI Design toolkit \citep{nokia} are examples of industrial contributions. These examples illustrate the variety of approaches used when designing RAI toolkits for both content and delivery medium. \textit{In terms of content}, while both MIT's AI Blindspot cards and Microsoft's Judgement Call are decks of cards, their content serves different purposes. Microsoft's cards aim to foster empathy in practitioners through gamefication, whereas MIT's cards aim to educate practitioners on how to identify and address potential blindspots while building AI systems by providing examples, recommendations, and stakeholders to engage with for each blindspot. \textit{In terms of delivery mediums}, Stanford's toolkit comes in the form of a collection of worksheets, templates and resources while Nokia AI Design toolkit comes in the form of an interactive website dedicated to a single tool to aid practitioners in ensuring they have made all the necessary considerations to build responsible AI.

By just considering these four toolkits, it already becomes apparent that RAI toolkits come in a variety of shapes and sizes, which have been recently comprehensively reviewed by Wong et al.~\cite{wong2022}. In terms of their presentation or display medium, these range from physical mediums, such as decks of cards \citep{MIT} and canvases \citep{canvas}, to digital mediums, such as code packages \citep{lime} or interactive websites \citep{nokia}. While physical mediums, and decks of cards especially, are heavily used across design fields \citep{aarts2020, roy2019, peters2021}, digital mediums provide the added benefit of interactivity and adaptability. Currently, there has been no exploration of the effects of RAI toolkits' medium or form on its users and the outcomes they produce.

\subsubsection{Design Decisions}
Nevertheless, not all RAI toolkits are equivalent or interchangeable. While still largely unexplored, the design decisions made when creating a RAI toolkit can impact its effect on those using it and the outcomes it helps to produce. For example, when working with non-technical stakeholders, a toolkit's use of metaphors to explain AI capabilities (e.g., anthropomorphise conversational AI \citep{luria2018}) has been found to be much more effective than simply listing capabilities \citep{lockton2022}. Another recently explored design decision is whether or not a toolkit de-couples or ``decontextualises" \citep{wong2022} its view of ethics from that of a specific domain or context. Such a design decision can significantly impact the use a toolkit by allowing practitioners to ignore contexts or abstract away inconvenient details, which, in turn, can encourage destructive behaviours such as shifting responsibility to other stakeholders \citep{morley2021}.

Wong et al.'s review of 27 toolkits that focus on AI ethics also highlights broader trends within the design decisions of these interventions \cite{wong2022}. In terms of narrative, toolkits tend to focus on either societal harms of AI or organizational risks. They also sometimes focus on `opportunities' as potential positive outcomes. Toolkits are either based on what is seen as responsible; on laws and regulations; or on some form of human rights, values or principles \citep{wong2022}. When aimed at developers and technical stakeholders, ethics is framed as a series of specifications or requirements, when aimed at business owners and executives, ethics is framed as business strategy and risk assessment. In terms of limitations, many toolkits focus on the technical aspects of ethics and make it difficult for non-technical stakeholders to get involved by offering little support for the ``translational work" needed to bridge between disciplinary knowledge \citep{wong2022, dolata2019}. They also advocate for stakeholder participation, but offer little guidance in terms of identifying and engaging stakeholders. 

Wong et al.~\cite{wong2022}'s work begins to provide a taxonomy for RAI toolkits based on certain design decisions, such as the narrative they support or the stakeholders they target. Our work is specifically interested in exploring which design decisions relate to the operationalisation of VSD's core principles and in what ways.

\section{Methodology}
\label{sec:methodology}

\textbf{\textit{Author Positionality Statement.}} To position ourselves as researchers and clarify our perspectives on the study \citep{frluckaj2022, havens2020}: This research was conducted in a Western, European context. The research team includes two women and two men from North-Eastern Africa, and Southern and Western Europe, working in academia and industry. With individual backgrounds in Human-Computer Interaction, Design, Computer Science, and AI, the team shares a common interest in the Design of Human-centred AI and Responsible AI. 

Below, we start by outlining the process used to select RAI toolkits for our user study (\S\ref{subsec:toolkits_selection}). This is followed by a discussion on how we mapped VSD values to those of RAI (\S\ref{subsec:mapping}), answering our (RQ\textsubscript{1}) and then describe how we conducted workshops to investigate whether and how RAI toolkits support VSD in their design and facilitate collaboration and learning (\S\ref{subsec:workshops}), answering our (RQ\textsubscript{2}).

\subsection{Selecting and presenting RAI toolkits}
\label{subsec:toolkits_selection}
% \subsubsection{Toolkit Selection}
To identify and select RAI toolkits for the workshops, we followed a similar methodology to that conducted by a recent review of RAI toolkits \citep{wong2022}. For our initial corpus, we reviewed a total of 63 toolkits; 27 toolkits from a recent RAI toolkits taxonomy~\citep{wong2022}, and an additional set of 36 toolkits from a large collection of practical tools for legal, ethical, and societal aspects of AI and data driven applications \citep{data-toolkits}. We removed two toolkits because they were duplicates in both sources (i.e., the AI Ethics Cards and Aequitas). Additionally, since the review of RAI toolkits was conducted in 2022~\cite{wong2022} and the online repository was undated, we included two additional toolkits that were released in 2023. Table~\ref{tab:toolkits-reviewed} shows the full list of toolkits that were reviewed, and Figure~\ref{fig:toolkits-considered} provides a breakdown of the selection process. The four-step process included:

\begin{itemize}
    \item \textbf{Step 1 - Target Users}: We selected toolkits designed for AI technical practitioners (e.g. developers, data scientists), resulting in 38 candidate toolkits.
    \item \textbf{Step 2 - Focus on Regulation}:  We excluded toolkits from regulatory institutes focusing on regulations, resulting in 21 candidate toolkits.
    \item \textbf{Step 3 - Indication of Use}: We excluded toolkits lacking evidence of recent use, resulting in 6 candidate toolkits. We did so by following Wong et al.~\cite{wong2022} methodology using proxies such as toolkits' appearance in practitioner-made resource lists, search rankings, and, signs of community use.
    \item \textbf{Step 4 - Comparability}: We selected toolkits with comparable design features, collaboration and learning support (i.e. content division, graphics or illustrations, and provocative cues/questions) in order to control for any effects on the study results. The resulting toolkits were Nokia AI Design toolkit and the MIT's AI Blindspots toolkit. For brevity, we will refer to them as the Nokia AI Design toolkit and the MIT Blindspots toolkit respectively.
\end{itemize}

\begin{figure*}[t!]
  \centering
  \includegraphics[width=\linewidth]{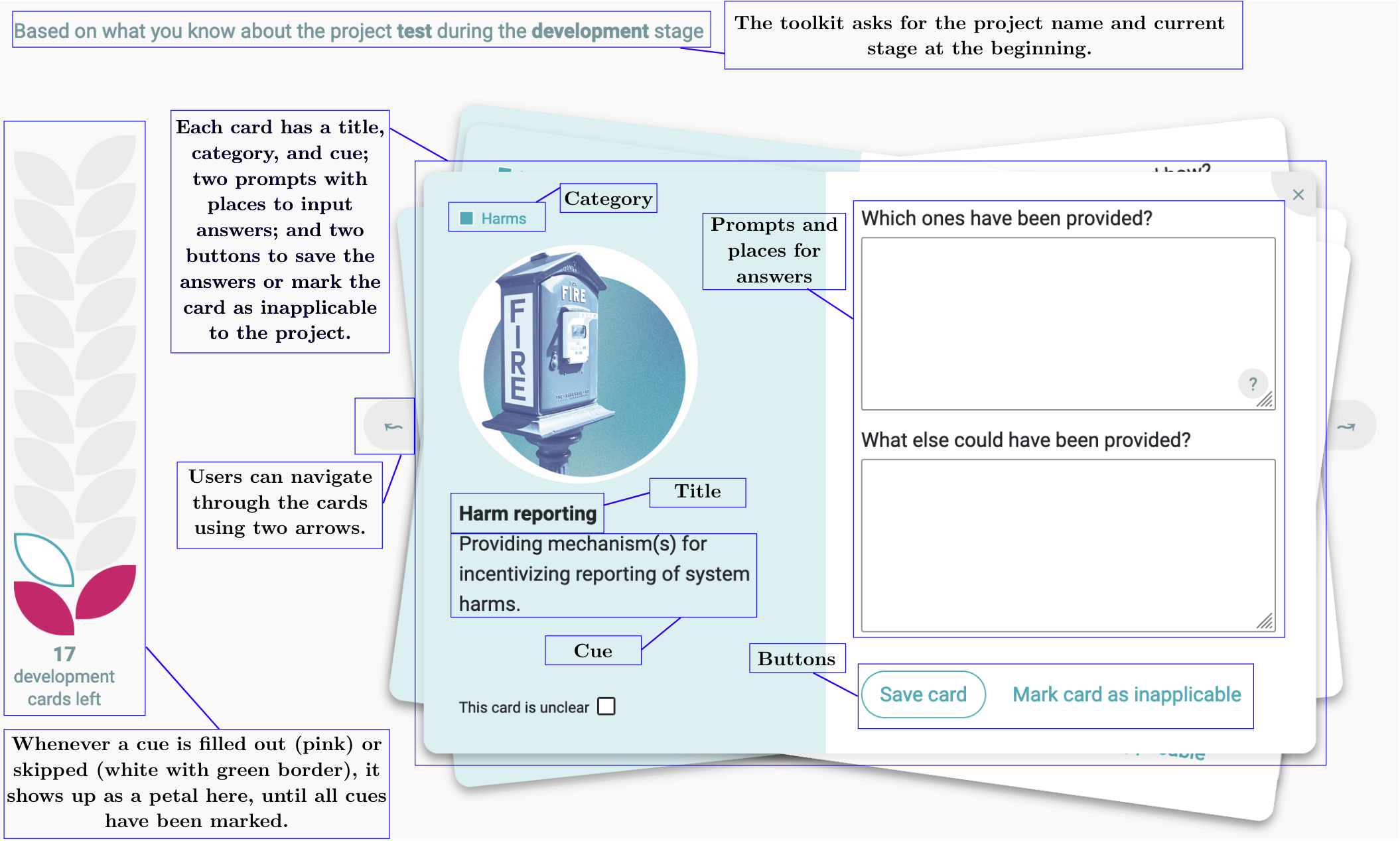}
  \caption{A screenshot of the Nokia AI Design toolkit with descriptions of each element in the interface in blue boxes.}
  \label{fig:toolkit-A-screenshot}
  \Description{A screenshot of the Nokia AI Design toolkit showing the interface of one of its cues. The screen shows a progress bar on the left; a card with a title, category, and cue; and text boxes to input answers to the prompts `which ones have been provided?' and `what else could have been provided?'}
\end{figure*}

\begin{figure*}[h]
  \centering
  \begin{subfigure}[]{0.7\linewidth}
         \centering
         \includegraphics[width=1.05\linewidth]{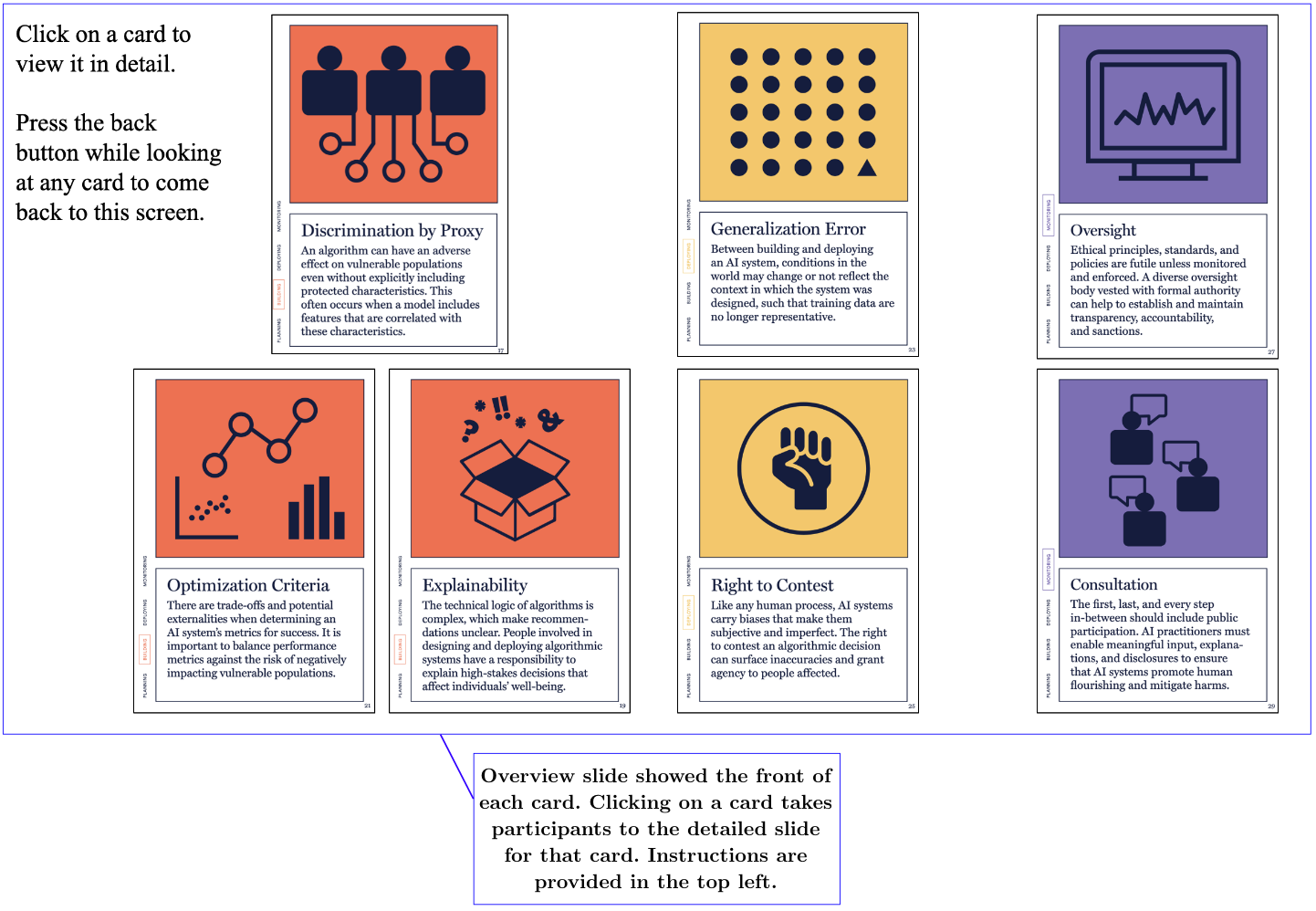}
         \caption{Screenshot of the MIT Blindspots toolkit's overview slide showing all the cards available. }
         \label{fig:toolkit-B-overview}
     \end{subfigure}
     \begin{subfigure}[]{0.7\linewidth}
         \centering
         \includegraphics[width=1.05\linewidth]{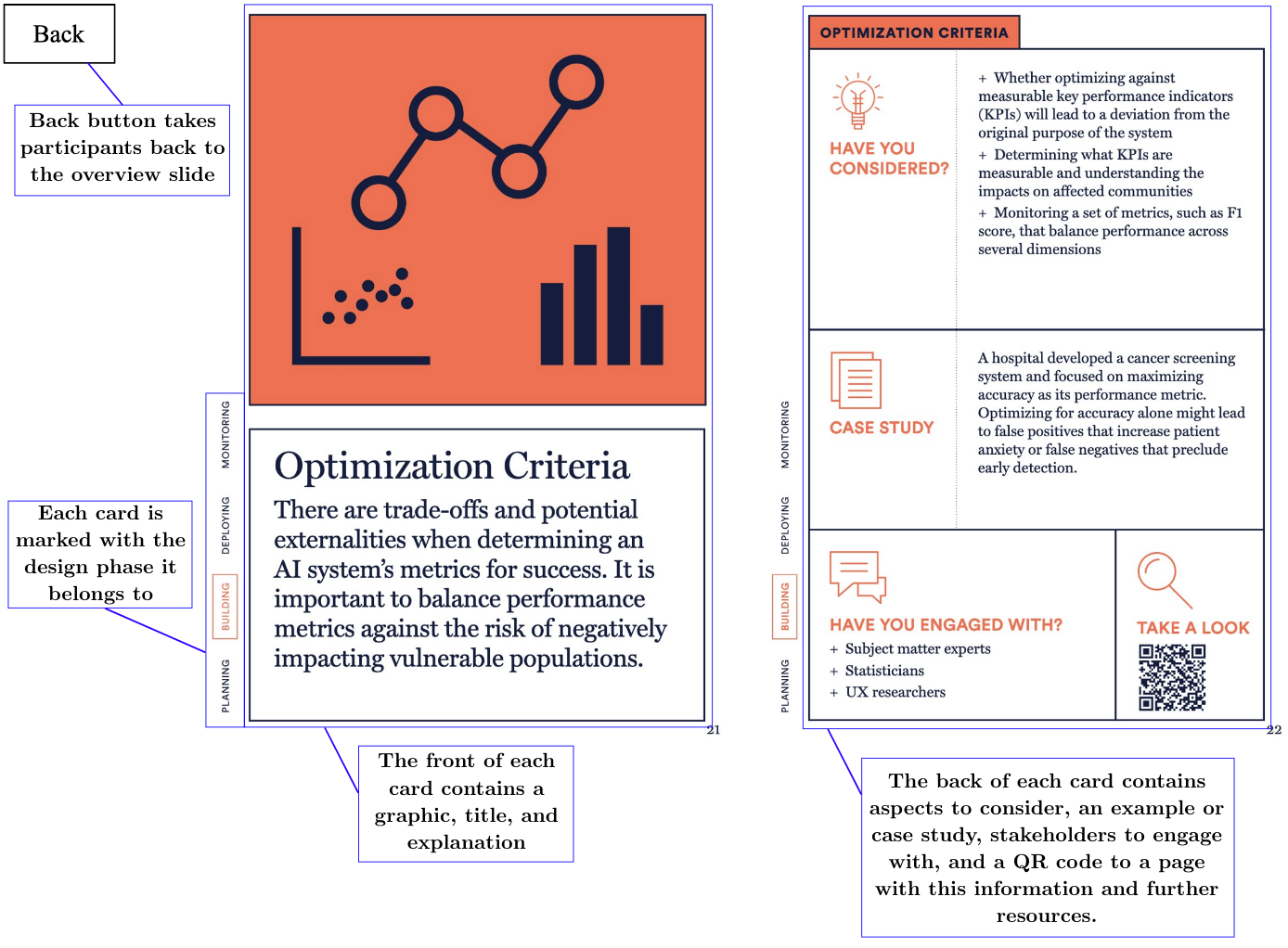}
         \caption{Screenshot of one of the MIT Blindspots toolkit's card slides showing one card in detail. }
         \label{fig:toolkit-B-detailed}
     \end{subfigure}
  \caption{Screenshots of the two types of slides in the MIT Blindspots toolkit: the overview slide (top), and an example of a detailed slide (bottom).}
  \label{fig:toolkit-B-screenshot}
  \Description{A screenshot of the MIT AI Blindspots toolkit showing the home view of all the 7 cards included, and a detailed view of a specific card showing its front and back.}
\end{figure*}

After selection, the toolkits were accessed and presented in the following manner for our study: \textbf{Nokia AI Design toolkit (Figure \ref{fig:toolkit-A-screenshot}):} We could access the source code, allowing us to create a copy without creator mentions. User interaction involved sequential card navigation with answer boxes, a progress bar, and an option to save and export answers. \textbf{MIT Blindspots toolkit (Figure \ref{fig:toolkit-B-screenshot}):} Unable to access the source code, we replicated the toolkit through a PowerPoint presentation. User interaction featured clickable thumbnails for detailed views, with QR codes on cards linking to additional information. 

\subsection{Mapping VSD values to RAI values}
\label{subsec:mapping}
Three of the authors went through the list of VSD values and those of RAI.
VSD defines a list of ``human values often implicated in system design" \citep{friedman2002}~[p.17] to consider (Figure \ref{fig:vsd-values}). Similarly, Responsible AI is about creating AI systems that are fair, transparent, and accountable, making a positive impact to society. To obtain the RAI values that are often used to design RAI toolkits, we relied on the NIST AI Risk Management Framework~\cite{nist_framework}. The framework identifies characteristics that contribute to AI systems that are fairness, explainable, accountable, privacy-preserving, secure and reliable, and sustainable. Alternatives include the Principled Artificial Intelligence from the Berkman Klein Center~\cite{fjeld2020principled}, which aligns with the NIST framework.

During this exercise, the authors found that it was difficult to conduct this mapping on the MIT Blindspots toolkit given the limited number of cards and due to the cards explicitly mentioning values such as fairness, explainability, accountability, safety, and so on; which defeated the purpose of the exercise. The Nokia AI Design toolkit proved much more effective due to the larger number and variety of cards, and the more implicit embedding of values within its cards and recommendations. 

\subsection{Conducting Workshops}
\label{subsec:workshops}

\begin{figure*}[t!]
  \centering
  \includegraphics[width=\linewidth]{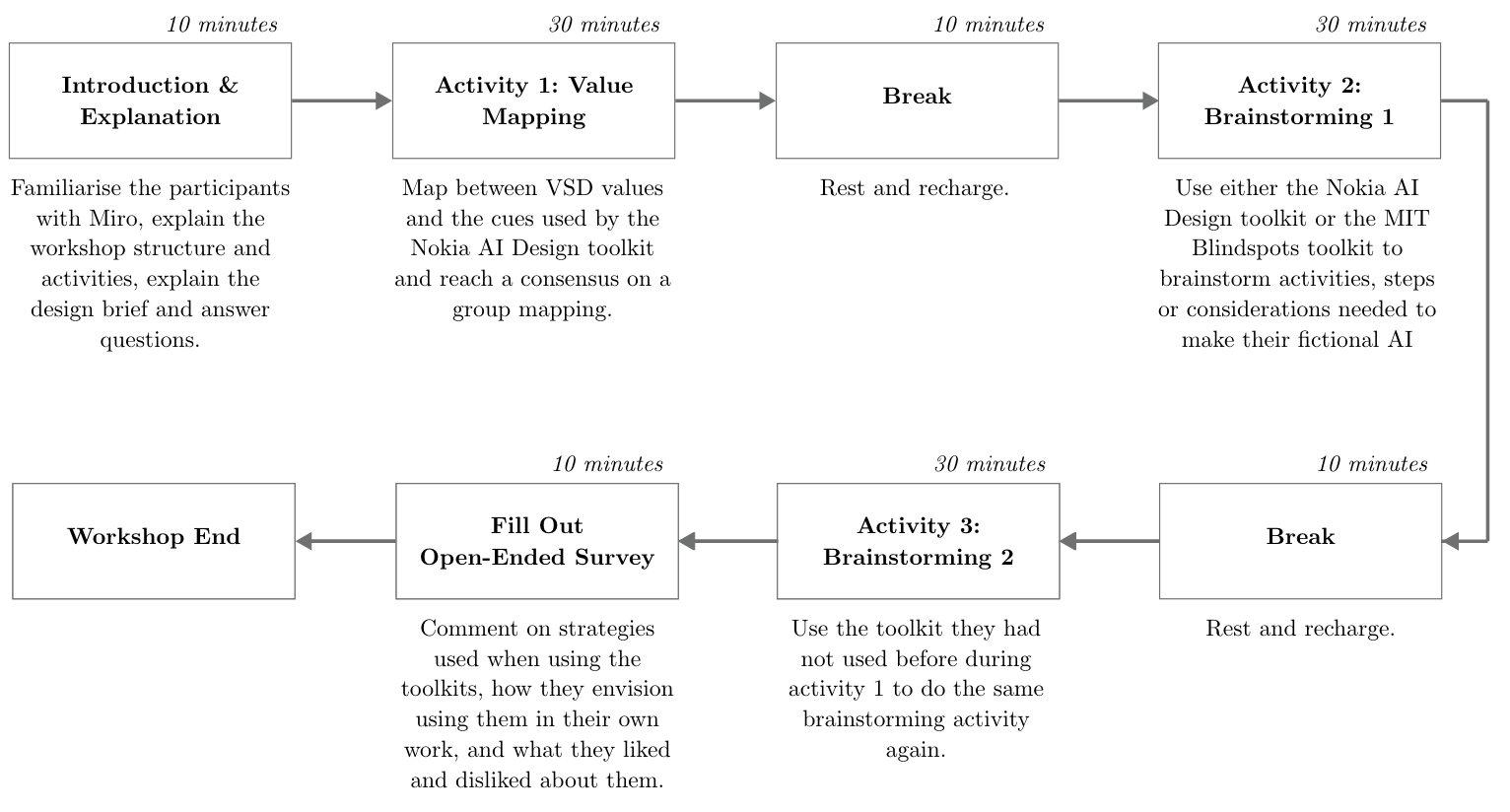}
  \caption{A flow diagram outlining the workshops' activities and their goals and duration. }
  \label{fig:workshop-structure}
  \Description{A flow diagram showing the different activities that took place during the workshop, along with their goals and duration. There were three 30-minute activities, with 10-minute breaks between each activity, and ending with an open-ended survey.}
\end{figure*}

The objective of our study was to investigate the extent to which VSD values align with RAI values, and whether and how RAI toolkits support VSD values in their design by promoting collaboration and learning. To do so, we conducted workshops with participants who engaged in value mapping and brainstorming while using selected RAI toolkits (Figure~\ref{fig:workshop-structure}). The use of collaborative design workshops has been recommended when creating responsible AI \citep{harbers2022} and is an effective approach for gathering interdisciplinary and in-depth insights \citep{martin2012} among several other benefits in the context of AI design \citep{sadek2023DS}. We first describe the participants, followed up by the workshop activities, then by the data collection and data analysis process.

\subsubsection{Participants}
Participants were recruited on a voluntary basis. The inclusion criterion was that participants were "familiar with how AI systems work" and "how to build at least one type of AI systems". This was checked through a questionnaire where participants described in details how they had learned these skills, e.g. through formal education (e.g., university courses) or self-learning (e.g., online courses). Participants were recruited throughout the study until a saturation was perceived to be reached (i.e., our process of interpreting the data collected yeilded no new insights \citep{braun2021}), in line with the grounded theory approach \citep{khan2014} and studies with similar methods \citep{sanderson2023}. Recognizing the subjective nature of saturation, in this study, saturation was deemed reached when the majority of themes generated post-workshop were consistent with themes identified in previous workshops \citep{braun2021}.

We recruited 17 participants (11 male, 6 female), whose ages ranged from 22 to 32 (M = 25.9, SD = 3.7). All were researchers with varying levels of experience with AI systems, ranging from 0.5 to 6 years (M = 2.9, SD = 1.6). Early-career researchers were screened to confirm either their current status as AI practitioners or their intention to pursue careers as AI practitioners. Table~\ref{tab:demographics} summarises participants' demographics. The study was approved by the Science Engineering Technology Research Ethics Committee at Imperial College London under the SETREC reference 21IC7361. Participants signed consent forms prior to attending the workshop, and received £25 Amazon gift cards as compensation for their involvement.

\begin{table}
    \centering
    \caption{Workshop participants' demographics.}
    \label{tab:demographics}

    \begin{tabular}{lllll}
    \toprule
    \begin{tabular}[c]{@{}l@{}}\textbf{Participant} \\ \textbf{ID}\end{tabular} & \textbf{Age} & \textbf{Gender} & \textbf{Role} & \begin{tabular}[c]{@{}l@{}}\textbf{AI} \\ \textbf{Experience} \\ \textbf{(years)}\end{tabular}  \\ \midrule
     1 & 32 & Male & MSc Student & 0.5 \\ 
    2 & 31 & Female & PhD Student & 1 \\ 
    3 & 22 & Male & MSc Student & 1.5 \\ 
    4 & 23 & Male & MSc Student & 2 \\ 
    5 & 26 & Male & MSc Student & 2 \\ 
    6 & 22 & Female & BSc Student & 2 \\ 
    7 & 26 & Male & PhD Student & 2 \\ 
    8 & 28 & Female & MSc Student & 3 \\ 
    9 & 22 & Male & MSc Student & 3 \\ 
    10 & 22 & Male & MSc Student & 3 \\
    11 & 22 & Male & MSc Student & 3 \\ 
    12 & 25 & Male & MSc Student & 3 \\ 
    13 & 29 & Female & PhD Student & 3 \\
    14 & 30 & Female & PhD Student & 4 \\
    15 & 22 & Female & MSc Student & 4 \\ 
    16 & 31 & Male & Post-Doc & 6 \\ 
    17 & 27 & Male & MSc Student & 7 \\     
    \bottomrule
    \end{tabular}%
\end{table}

\subsubsection{Workshop Structure and Activities}
The workshop was 2.5 hours long, and was conducted 4 times in Spring 2023, with different participants each time (groups of 4, 3, 4, and 6). It was conducted online using Microsoft Teams and Miro, and consisted of three activities and three surveys (following each activity). The overall procedure for the study is shown in Figure \ref{fig:workshop-structure} and described below.

\begin{figure*}[t!]
  \centering
  \includegraphics[width=\linewidth]{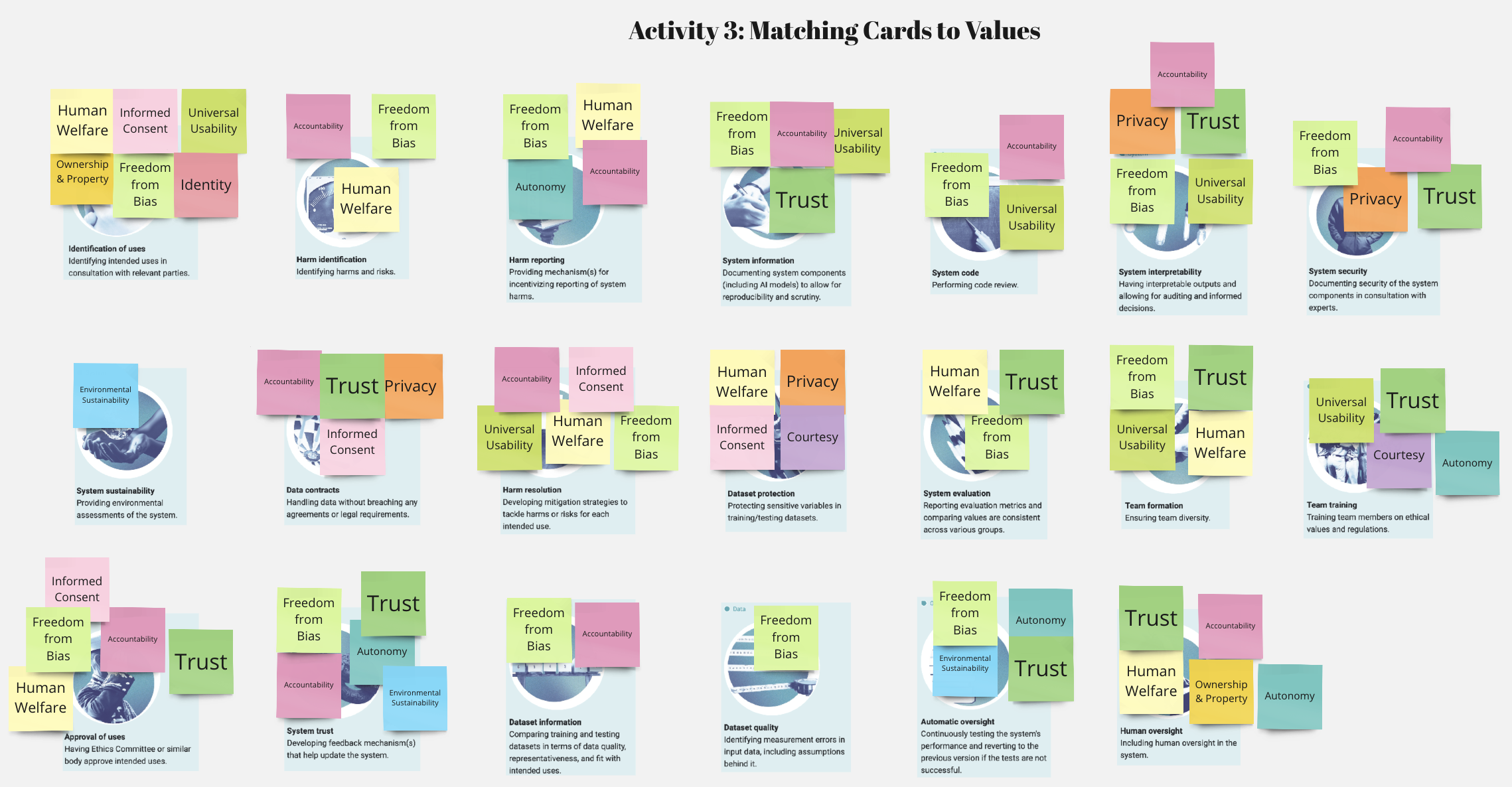}
  \caption{The Miro board for Activity 1 after participants had assigned values to the Nokia AI Design toolkit and reached a consensus.}
  \label{fig:activity-3}
  \Description{A screenshot of the Miro board after Activity 1 was conducted and participants had assigned values to different Nokia AI Design toolkit cues using sticky notes.}
\end{figure*}

\smallskip
\noindent\textbf{Activity 1 (Value Mapping)} lasted 30 minutes. The goal of this activity was to empirically obtain a mapping between the Nokia AI Design toolkit and VSD values. The structure of this activity was derived from the methodological approaches of affinity mapping \citep{harboe2015} and card sorting\citep{hawala2006} as methods of directly assigning values by participants, as opposed to more subjective and implicit methods used in previous works \citep{stoimenova2020}. First, participants were asked to re-read the cards in the Nokia AI Design toolkit and then read a list of ``universal values'' outlined by VSD as being ``often impacted upon by technology'' \citep{friedman2006}. The Miro board layout for this activity is shown in Figure \ref{fig:activity-3}. Participants were then asked to assign values to each card based on which values they felt the card was respecting or advocating for. They were told that they could assign multiple values to each card and assign a value to multiple cards. This was done individually for 15 minutes and then as a group for 15 minutes where participants aggregated all the values they had assigned to each card and then discussed and changed values until a consensus was reached for each card. Participants were then given another 15 minutes to go through the values assigned by the whole group, discuss any discrepancies, and reach a consensus together.

\smallskip
\noindent\textbf{Activities 2 and 3 (Brainstorming)} lasted 30 minutes each. The goal was to learn about how participants used and envisioned themselves using the toolkits, and to analyse the differences between the two toolkits in terms of their effects on participants and the ideas they helped them produce. Participants were given access to the toolkits through links embedded in the Miro board. To increase the comparability between the toolkits, both were presented in an interactive form, and we only included cards that related to the same phases (i.e. ``designing", ``deploying" and ``using" in the Nokia AI Design toolkit; ``building'', ``deploying'', or ``monitoring'' in the MIT Blindspots toolkit) . As such, the Nokia AI Design toolkit had 20 cards and the MIT Blindspots toolkit had 7 cards. 

\bigskip
\noindent Participants were either assigned the Nokia AI Design toolkit for Brainstorming 1 then MIT Blindspots toolkit for Brainstorming 2 (N = 8) , or the MIT Blindspots toolkit then the Nokia AI Design toolkit (N = 9). Both these activities were conducted by participants individually. 
Participants were asked to brainstorm as many activities, steps, or considerations needed to ensure that a fictional AI system is `responsible' before deployment. They were given the following fictional scenario to work with:

\begin{quote}
    \emph{``You are on a team building an AI-powered chatbot for your company that will help people self-manage their health. The initial planning and design phases are complete and you are now building and training the AI model. You need to make a list of activities, steps, or considerations that your team will need to make moving forward to ensure the chatbot is responsible and ethical. These should be focused on the building, deployment and monitoring phases."}
\end{quote}

Healthcare-related use-cases have been used in previous studies when exploring aspects relating to responsible AI \citep{liao2022}. This speculative healthcare context was chosen as a context that many participants are likely to be familiar with, and as suitable context to explore human values \citep{strikwerda2022}. It is not our intent to focus this work solely on AI for healthcare or frame our contributions as such. Furthermore, the addition of the AI-powered chatbot was made to provide a relatable, relevant and interesting AI technology given the recent advent of large language models such as the ChatGPT. 

\smallskip

\begin{table*}[t!]
    \centering
    \caption{Mappings between the Nokia AI Design toolkit's cues, their responsible AI pillars, and the VSD values assigned to them by consensus.}
    \label{tab:mapping}
    \Description{A table showing the cues included in the Nokia AI design toolkit along with their corresponding RAI value and the the VSD values assigned to them by consensus.}
    \resizebox{\textwidth}{!}{%
    \begin{tabular}{lll}
    \toprule
    \textbf{Nokia AI Design Toolkit} & \textbf{RAI Value} & \textbf{VSD Value} \\ \midrule
    Identifying intended uses in consultation with relevant parties & Accountability & No consensus \\
    Having an ethics committee or similar body approve intended uses & Accountability, Safety & Human Welfare \\ 
    Providing mechanism(s) for incentivising reporting of system harms & Accountability & Accountability\\ 
    Developing mitigation strategies to tackle harms or risks for each intended use & Fairness & Freedom from Bias\\ 
    Documenting system components (including AI models) to allow for reproducibility and scrutiny & Accountability, Fairness & Accountability \\ 
    Performing code review & Accountability, Fairness, Reliability & Accountability\\ 
    Reporting evaluation metrics and checking them in different groups & Fairness, Transparency & Trust\\ 
    Having interpretable outputs to allow for auditing purposes and informed decisions &  Transparency & Trust \\ 
    Documenting security of the system components in consultation with experts & Security & Privacy, Trust\\ 
    Providing environmental assessments of the system & Sustainability & Environmental Sustainability\\ 
    Developing feedback mechanism(s) that help update the system & Accountability & Accountability\\ 
    Handling data without breaching any agreements or legal requirements & Accountability & Accountability\\ 
    Comparing training and testing datasets in terms of data quality, representativeness, and fit with intended uses & Fairness, Transparency & Freedom from Bias\\ 
    Identifying measurement errors in input data, including assumptions behind it & Transparency & Trust\\ 
    Protecting sensitive variables in training/testing datasets & Accountability, Privacy & Privacy\\ 
    Continuously testing the system's performance and reverting to the previous version if the tests are not successful & Reliability, Safety, Security & Trust\\ 
    Including human oversight in the system & Accountability & Accountability \\ 
    Ensuring team diversity & Accountability, Fairness & Freedom from Bias\\ 
    Training team members on ethical values and regulations & Accountability & No consensus\\ 
    \bottomrule
    \end{tabular}%
    }
\end{table*}

\subsubsection{Data Collection \& Analysis}
The study employed a qualitative approach using data collected throughout the workshops: value mappings, transcripts and outcomes of brainstorming activities, and participants' responses to the open-ended survey.

Thematic analysis \citep{braun2012} was used to identify and cluster themes the researchers' identified within activity outcomes, workshop transcripts and open-ended survey questions. Initially, top-down coding relied on researchers’ workshop observations (e.g ``mentioning examples”) and based on conceptual categories (e.g. ``negative aspects mentioned regarding the Nokia AI Design toolkit/the MIT Blindspots toolkit"), while subsequent bottom-up coding constructed sub-themes based on researchers’ understanding of the data \citep{braun2019}. The analysis, conducted in Miro using sticky notes from the workshop, involved participants’ survey answers, quotes, and researchers’ observations from the workshops and transcripts. These sticky notes were then clustered by the researchers into the top-down themes mentioned earlier. Afterwards, individual researchers organised themes into sub-themes using a bottom-up approach. Finally, discussions took place until a consensus was reached.

The resulting themes (see 4.2.2 in Section \ref{sec:results} for details) are as follows: ``navigation", ``considering stakeholder perspectives", ``collaboration versus solo work", ``open-ended cuing", ``user experience and content", ``lack of adaptive responses", ``providing examples and case studies", ``practical support needed".
\section{Results}
\label{sec:results}

\subsection{\textbf{RQ\textsubscript{1}:} How closely do Value Sensitive Design (VSD) values align with Responsible AI (RAI) values integrated into RAI toolkits?}

\label{sec: RQ1}

VSD values align, to a great extent, with RAI values. When asked to map between the Nokia AI Design toolkit and VSD values, participants considered a number of stakeholders' perspectives and considered the values from the point of view of testers, developers, ethics boards, users. Overall, a consensus was reached between the experts' mapping and the workshop participants' mappings on all but three cards for: `identifying intended users in consultation with relevant parties', `training team members on ethical values and considerations', and `having an ethics committee or similar body approve of intended uses' where the researchers assigned all three the value of `accountability' and workshop participants assigned them the values of `informed consent', `universal usability', and `human welfare' respectively. This has lead to an overall consensus of 85\% (17/20 cards) across researchers and workshop participants for the following mappings:

Overall, the value of accountability was most implicitly represented by the cards provided within the Nokia AI Design toolkit (6/20), followed by trust (5/20). Out of the 13 VSD values provided by \citep{friedman2002}, 6 are represented in the toolkit's cards, although all the VSD values were assigned to various cards by at least one person during the workshops. The fact that the three cards where a consensus was not reached were assigned the value of `accountability' by researchers indicates that the conceptual definition for that value held by the researchers might have differed from participants, which is supportive of previous work highlighting different groups having different value definitions and priorities \citep{jakesch2022}.

It was interesting to note that almost all workshop participants struggled with the values of `calmness' and `courtesy' as they felt ``unfamiliar" with them and were ``not how they would refer to AI ethics aspects". One participant mentioned that an alternative value to those could be ``competence or effectiveness" in the sense of ``acting with due diligence, care and vigilance and making sure quality was good enough". Four participants also felt that these values were ``secondary byproducts" as opposed to ``primary concerns" for them. They appreciated that the cards would ``factor in" or consider these values for them in the actions and recommendations they offered so that they would not have to think about them actively themselves. 

\noindent\textbf{Summary.} A consensus was reached between the experts' mapping and the workshop participants' mappings on all but three cards where the researchers had assigned the value of `accountability'. The cards represented the values of `accountability' and `trust' most commonly. Practitioners struggled with unfamiliar values and felt that some values had a secondary importance.

\label{sec:RQ2}

\subsection{\textbf{RQ\textsubscript{2}:} How do existing RAI toolkits incorporate VSD, and support collaboration and learning?}

\smallskip

We begin by discussing the outcomes of the brainstorming session, followed by the design choices of the two toolkits and their support for collaboration and learning. 

\begin{figure*}[t!]
\centering
\begin{subfigure}[b]{0.8\textwidth}
   \includegraphics[width=1\linewidth]{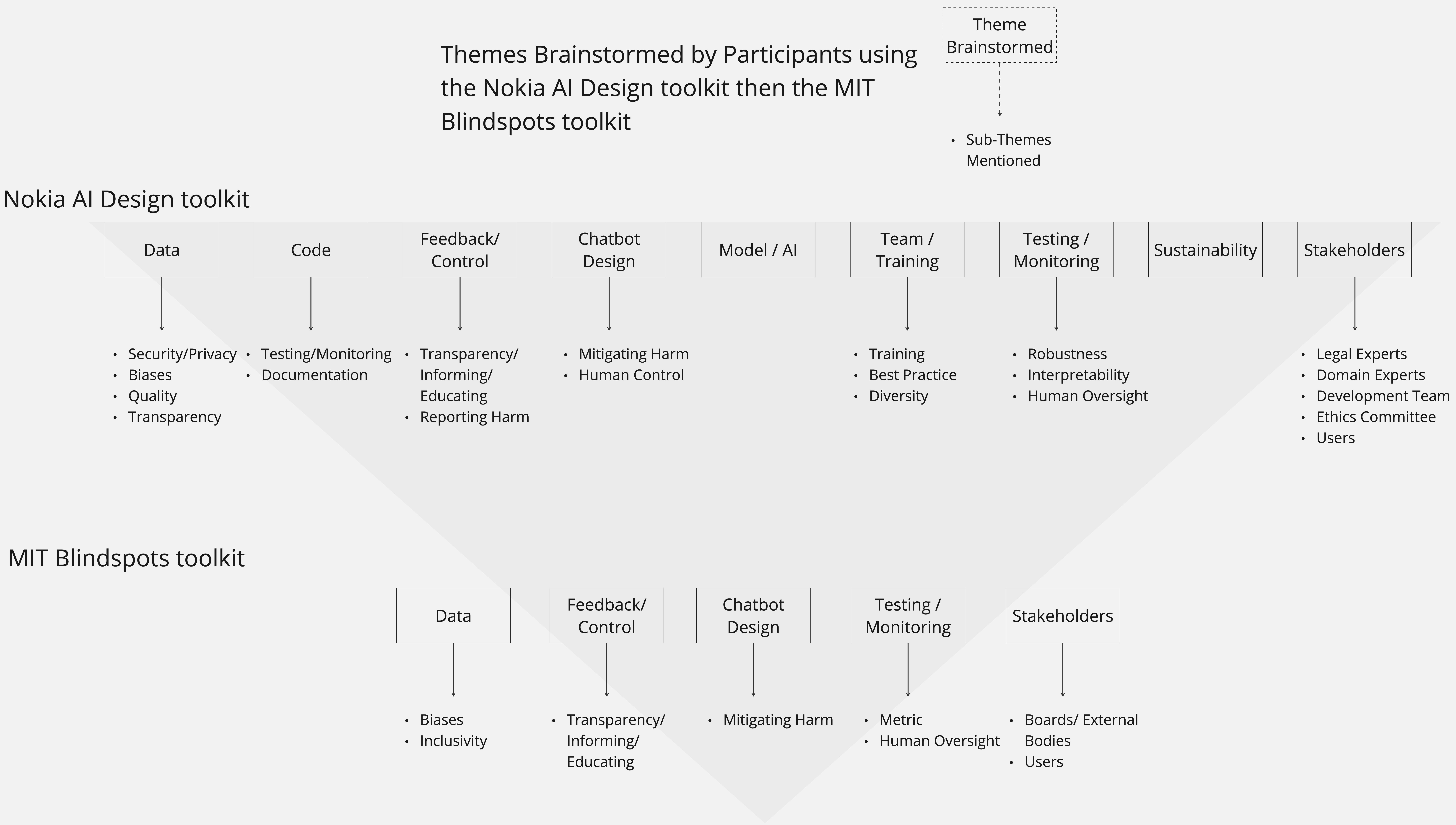}
   \caption{Ideas brainstormed by participants using the Nokia AI Design toolkit then the MIT Blindspots toolkit.}
   \label{fig:coding-tree-A} 
\end{subfigure}

\begin{subfigure}[b]{0.8\textwidth}
   \includegraphics[width=1\linewidth]{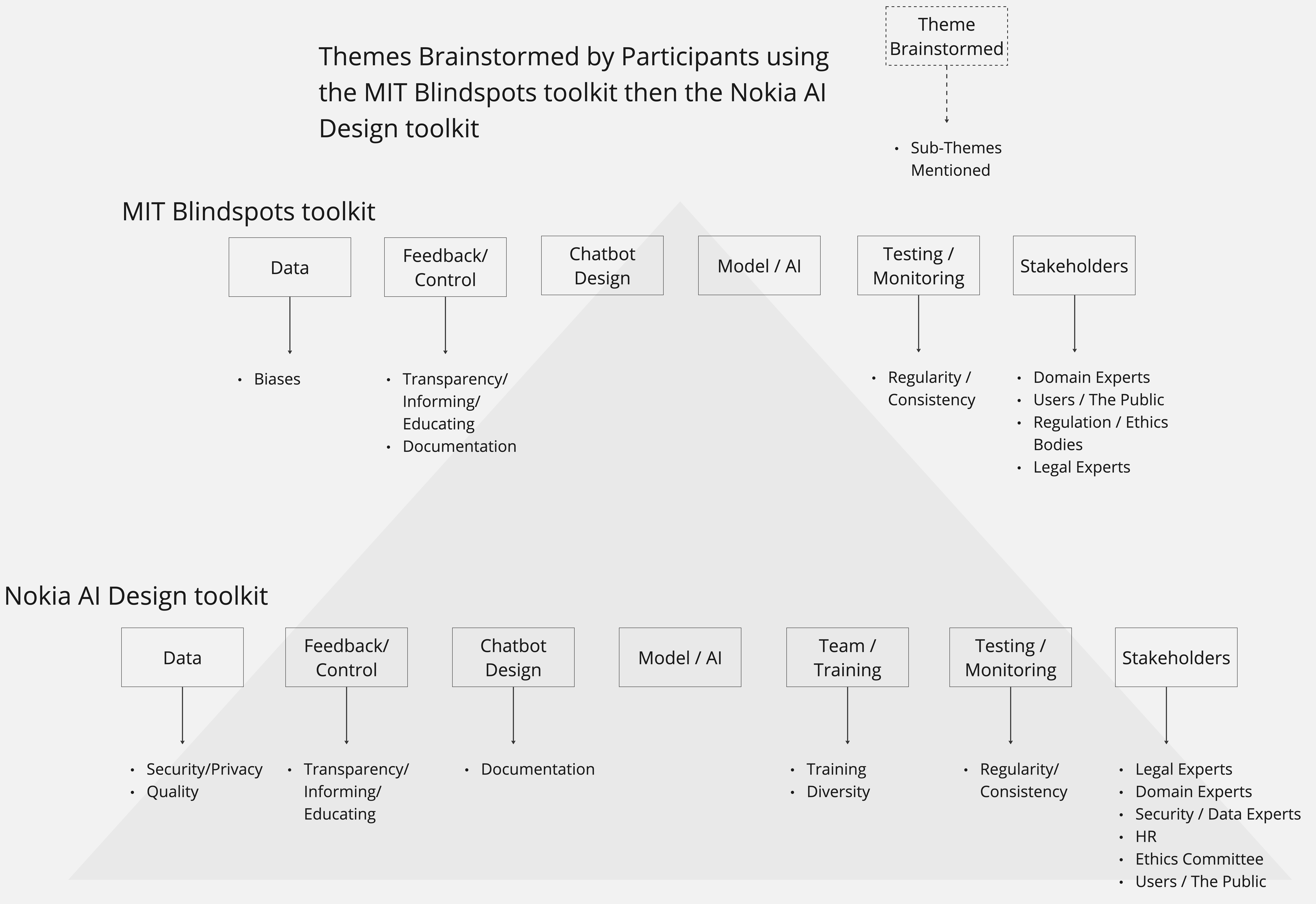}
   \caption{Ideas brainstormed by participants using the MIT Blindspots toolkit then the Nokia AI Design toolkit.}
   \label{fig:coding-tree-B}
\end{subfigure}
\caption {Coding trees for both toolkits. (a) Participants who used the Nokia AI Design toolkit then the MIT Blindspots toolkit; (b) Participants who used the MIT Blindspots toolkit then the Nokia AI Design toolkit. Each box represents a main theme, and is then broken down into sub-codes that represent the ideas under each theme that participants touched upon in their brainstorming sessions.}
\label{fig:coding-trees}
\Description{Two figures showing coding trees for the codes assigned to sticky notes created by participants during Activities 1 and 2. The top tree is for Participants who used the Nokia AI Design toolkit then the MIT Blindspots toolkit, and the bottom tree is for Participants who used the MIT Blindspots toolkit then the Nokia AI Design toolkit. Each box represents a main theme, and is then broken down into sub-codes that represent the ideas under each theme that participants touched upon in their brainstorming sessions.}
\end{figure*}

\smallskip
\noindent\textbf{Outcomes of Brainstorming Session.}
Participants who used the Nokia AI Design toolkit first generated a total of 84 ideas across the 4 workshops, and a total of 35 ideas when they then used the MIT Blindspots toolkit. Conversely, participants who used the MIT Blindspots toolkit first generated 42 ideas across the workshops and 69 ideas when they then used the Nokia AI Design toolkit In both cases, participants generated a higher number of ideas using the Nokia AI Design toolkit, even when starting with the MIT Blindspots toolkit, despite our expectation that participants' second activity might generate fewer ideas than their first. Participants using the Nokia AI Design toolkit also generated a greater breadth of ideas and had a greater range of considerations under each idea or theme. Figure \ref{fig:coding-trees} shows the coding trees for the themes generated while brainstorming using each toolkit in both orders to highlight a disparity across toolkits. 

\smallskip
\noindent\textbf{Design Choices.} In terms of navigation, participants contrasted the navigation strategies that both toolkits afforded. On one hand, the MIT Blindspots toolkit offered more back-and-forth navigation. Three participants preferred being able to return to the `overview' screen and select the desired card. On the other hand, the Nokia AI Design toolkit offered more sequential navigation. Four participants preferred this more sequential nature as it forced them to consider each card \emph{``one by one"} and write down \emph{``what ideas or actions it made [them] consider"} and not skip ones they assumed to be irrelevant.

Four of the participants who started with the Nokia AI Design toolkit filled out their answers directly into the tool itself and sent the generated PDF to the researchers as opposed to using the Miro board, and enjoyed using the interface directly. Participants also strongly appreciated the ability to save their responses as a PDF afterwards. One participant mentioned that the \emph{``PDF consolidated review and the option to upload an old result for comparison/review was really good"} and another participant complained that in the MIT Blindspots toolkit there was \emph{``no way to evaluate or summarise [their] ideas/thoughts as [they] go through the toolkit"}.

Participants also commented on the Nokia AI Design toolkit's flexibility, allowing them to use it throughout the design process and therefore making them feel more efficient and productive during the brainstorming activity. Because of its ability to diverge during brainstorming sessions, four participants felt that they would want to use the Nokia AI Design toolkit during early planning phases of a project as a \emph{``starting point"} to \emph{``devising a plan on how to design an AI"}, \emph{``make [them] think about what [they] would need to think about to ensure this tool is ethical"},  and \emph{``adjust the system design to be more responsible and ethical"}, as well as to \emph{``raise [their] concerns to [their] colleagues and team and communicate [their] views."} Five participants felt that they could also use the toolkit towards the end of a project for \emph{``auditing", ``testing", and ``evaluation",} with two participants stating that it could be used repeatedly throughout. Participants described their brainstorming sessions as \emph{``more productive", ``more aware", ``more critical", ``more efficient", and ``more comprehensive"} having used the Nokia AI Design toolkit. Four participants referred to the toolkit giving them ideas that \emph{``did not come to mind",} and \emph{``new solutions"} that \emph{``cover blindspots"} they originally had.

Participants also commented about the Nokia AI Design toolkit's lack of adaptive responses. One participant felt they could fill in anything in the boxes provided and the tool would say \emph{``well done"} or \emph{``40\% done/considered"}, but that would not be true and would be misleading to think. Another participant felt that the Nokia AI Design toolkit \emph{``did not evaluate at all what [they] wrote"} and was \emph{``unusable",} and another commented that the tool \emph{``does not give an actual indicator of how good the system is already"}. They were worried that the tool relied too much on how well and how reliably people explain their systems. One participant suggested that the tool should \emph{``provide a more custom response (e.g., analyse the Github repo and answers) instead of just repeating user input".}
\smallskip

\noindent\textbf{Collaboration and Learning.} In terms of collaboration, participants found that the Nokia AI Design toolkit fostered more open-ended brainstorming and thus allowed for discussion and collaboration, especially within teams. They described their brainstorming with the toolkit as \emph{``organic"} and \emph{``unbiased"} given the lack of direction and the open/general nature of the cards and that the toolkit \emph{``supported open-ended ideation to let out what you feel"} and \emph{``provided many ideas from which it was easy to formulate more specific activities and considerations".} Participants used the cards more as starting points or \emph{``springboards"} that allowed them to ``sprout ideas" and provide \emph{``inspiration".} 

Six participants felt that the the Nokia AI Design toolkit was more suited for collaborations. One participant commented that \emph{``[the] toolkit is good in team meetings because it's more open-ended and people can contribute more"} and that it was useful to \emph{``get the ball rolling"}. Nine participants indicated that the MIT Blindspots toolkit was more suited for solo work whereas the Nokia AI Design toolkit was more suitable for team collaborations. They felt that the MIT Blindspots toolkit could be used as an \emph{``education tool"}, mainly because it \emph{``provides examples and some insights to people who might not be familiar with all the different aspects of RAI"}.

Participants discussed using the two toolkits together: \emph{``[the MIT Blindspots toolkit] offers more concrete instructions where you can go back and flesh out your ideas"} after diverging with the Nokia AI Design toolkit since \emph{``as one person you would need more cuing [than is available in the Nokia AI Design toolkit]"} to flesh out your ideas. Another participant also echoed this sentiment stating: \emph{``in an individual setting, I felt like it [the Nokia AI Design toolkit] needed more direction."} Finally, another participant mentioned that: \textit{``A good strategy would be to start with [the Nokia AI Design toolkit] to brainstorm and think about the issues, then use [the MIT Blindspots toolkit] to learn more about how to handle the issues, then go back to [the Nokia AI Design toolkit] to build on your ideas".}

Participants also commented about toolkits ability to help them consider different stakeholders' perspectives. Five participants reported that the Nokia AI Design toolkit encouraged them to consider aspects from the perspectives of different stakeholders including users, the development team, the legal team, and ethicists. They also mentioned perspectives in the sense of considering the system from a \emph{``system's, sustainability, usability, trust, and privacy perspective".} 

Finally, the key strength of the MIT Blindspots toolkit lies in its provision of examples and case studies, as well as recommendations and suggestions fostering learning. All study participants showed appreciation towards these aspects of the toolkit. One participant even felt that brainstorming would be redundant as they felt that all the considerations they needed to think of were provided by the MIT Blindspots toolkit's recommendations already. Participants felt that the case studies and examples offered ``directed action", whereas \emph{``it was difficult to apply [the Nokia AI Design toolkit]'s questions to a specific use-case"} and that \emph{``the cards are quite general and it required thinking about how this applied to our specific scenario"}. Another participant also stated that they had to read the Nokia AI Design toolkit's cards twice: once to understand them, then once to apply them to the workshop's specific scenario. Two participants also felt that if you did not understand one of the Nokia AI Design toolkit's cards, there were no examples to help. Finally, another participant also felt that there were too many assumptions they had to make to fit the Nokia AI Design toolkit to their application. Nonetheless, one participant also noted that the \emph{``case study was useful but probably pigeon-holed you into thinking about the problem in that dimension only"}. Regarding providing recommendations and solutions, two participants felt that with the Nokia AI Design toolkit if you do not know how to solve a specific issue it offers no guidance, while the MIT Blindspots toolkit had case scenarios and examples you could build upon. One participant also felt that \emph{``[the Nokia AI Design toolkit] just said `have you thought about this or that' but there was nothing about what to do about it"} and another participant built upon that stating: \emph{``I am ignorant [about legal aspects], and the cards did not provide any help except highlighting my ignorance [referring to the Nokia AI Design toolkit]".}

Nevertheless, few participants found the MIT Blindspots toolkit's recommendations too vague and overwhelming. They did not want a \emph{``you need to do this"} approach, but more practice-based help such as providing tool recommendations. One participant referenced Adobe tools showing you how to improve accessibility and another participant mentioned wanting a code analysis tool that flags problematic parts of the code as more helpful ways to provide support than the MIT Blindspots toolkit's recommendations. They mentioned that the suggestions felt very broad (e.g., documenting the security of all system components in consultation with experts) and almost like PhD projects (e.g., creating a tool to measure fairness). Interestingly, one participant mentioned that all the toolkit does is keep asking \emph{``have you considered this?"}, continuing on to reply \emph{``no, and I didn't have time to".} This highlights participants' negative attitude and feeling of being overwhelmed because of this lack of practical support.

\smallskip
\noindent \textbf{Summary.} Overall, during the brainstorming session, participants were able to generate a greater breadth of ideas and go into more depth with each idea using the Nokia AI Design toolkit. They also felt it allowed them to consider more stakeholders' perspectives. Participants also enjoyed its flexibility, the ability to fill out answers directly in the tool, and the ability to save their responses. Participants felt the Nokia AI Design toolkit was more suited for collaborative work, whereas the MIT Blindspots toolkit was better suited for solo work and education. The Nokia AI Design toolkit's open-ended cuing supported its use during different design phases and more divergent brainstorming. Finally, participants felt that the MIT Blindspots toolkit provided more information and found its provision of examples and case studies educational, but some participants felt that the recommendations provided were too vague and general.

\section{Discussion}
\label{sec:discussion}

By conducting four workshops with 17 AI researchers, we established that VSD and RAI values align to a great extent, and, as such, we explored the effects of toolkits' design features regarding collaboration and learning. We identified a number of links between toolkits' design choices, which resulted in differences in the ways participants perceived and interacted with them. First, our participants generally found the MIT Blindspots toolkit more suitable for individual work and the Nokia AI Design toolkit better for collaboration due to its generalisability and open-endedness. The Nokia AI Design toolkit facilitated broader ideation, evident in the quantity and variety of ideas generated and the breadth of categories in its coding trees. Conversely, the MIT Blindspots toolkit's provision of examples, case studies, solutions, and recommendations was a key discussion point during the workshops. In contrast, the absence of such elements in the Nokia AI Design toolkit required participants to engage more deeply and spend more time understanding its content. Finally, participants had mixed reactions to design features unrelated to content such as the order and number of cards, and navigation options. Non-linear navigation was appreciated for its flexibility, while a linear approach ensured thorough consideration of all cards.

Next, we synthesise these results into a number of theoretical implications in terms of links between the toolkits' design features and their support of VSD (\S\ref{subsec:theoretical_implications}), and practical implications revolving around the toolkits' ability to operationalise VSD by supporting collaboration and learning (\S\ref{subsec:practical_implications}).

\subsection{Theoretical Implications}
\label{subsec:theoretical_implications}

\smallskip 
\noindent\textbf{Discrepancies between RAI and VSD.} Overall, the RAI values matched the VSD values closely. However, our workshop participants and the study researchers did not reach consensus on two values: \emph{accountability} and \emph{transparency}. For accountability, the lack of consensus might be explained by recent empirical evidence illustrating that different groups of people define (and prioritize) responsible AI values differently~\cite{jakesch2022}. This suggests that seemingly similar sets of values should not be used interchangeably without a thorough understanding of their fundamental differences. For transparency, which VSD often refers to as trust, the picture was slightly different. While transparency and trust are certainly intertwined~\citep{vereschak2022}, transparency has been found to both enable and violate trust depending on contextual factors \citep{yu2022}. For example, revealing an AI model's low confidence score for its prediction might reduce trust in its competence, while increasing trust in its honesty. VSD has also been used as a facilitator for transparency \citep{dexe2020}, despite not including the value explicitly in the original set provided. Given the significance of transparency in AI systems \citep{lindley2020} and its established distinction from trust, we suggest that VSD's values ``often implicated in system design" \citep{friedman2002} should be updated to reflect these findings. While the VSD methodology based on conceptual, empirical, and technical investigations has been recently adapted to AI systems \citep{zhu2018, umbrello2021}, updating the core set of values included in the VSD framework has been largely remained unexplored.

\smallskip
\noindent\textbf{Framing RAI Toolkits in Research Taxonomies.} Wong et al. (2022) \cite{wong2022} discuss how different RAI toolkits frame ethics, and the discourse they use around ethical concepts. For example, some toolkits choose to focus on risks and negative outcomes, while others highlight the benefits and positive outcomes of building responsible AI. We propose adding a new dimension regarding the framing of the toolkits themselves (e.g., as educational, collaboration, or reflection tools). Framing the types of support that RAI toolkits offer, or the activities they can facilitate, can help their users to select appropriate toolkits more effectively, especially given the large number and variety available \citep{wong2022}. Our research shows that this framing is not always necessarily intended by the toolkit creators, it tends to be more implicit and depends on the design features of the toolkit. For example, participants felt that the Nokia AI Design toolkit's ability to support ``organic" brainstorming through its more open-ended cues was conductive of discussions and collaboration in team settings. Previous studies have also established the role of open-ended discussions in supporting collaboration \citep{mattelmaki2011}. On the other hand, the MIT Blindspots toolkit was perceived as an educational tool because of its detailed examples, case studies, and recommendations. 

\smallskip
\noindent\textbf{Low Actionability in RAI Toolkit Recommendations.} While previous work has shown that the use of examples and analogies can help establish empathy \citep{holden2018} and lower the effort needed during learning \citep{bolkan2019}, participants still felt that the MIT Blindspots toolkit's recommendations fell short of being practically meaningful. A large body of work has recently surfaced discussing the limitations of general recommendations for responsible AI \citep{harbers2022}. These works echo the sentiments of participants and also call for more practical tools and frameworks to improve actionability \citep{coldicott2019}. In this study, participants thought the recommendations provided were too general or too large in scope, making participants feel ``overwhelmed" and even ``ignorant". Participants' suggestions on how to improve these recommendations all revolved around analysing their code, providing links to specific tools that can be used, and giving more customised feedback. Similar to Wong et al. (2022) \citep{wong2022}'s finding that most RAI toolkits recommend involving stakeholders but offer no practical guidance on how to do that, this study also adds that this lack of practical guidance for advocated actions extends beyond involving stakeholders and generalises to a number of different recommendations provided by these toolkits.

\smallskip
\noindent\textbf{Considering the Role of Non-Content-Based Features of RAI Toolkits.} Participants were affected by a number of design decisions unrelated to the toolkits' content. Participants frequently separated their comments on toolkits' content (e.g., the cards given, ideas expressed, and text used), the presentation of that content and the interactions the toolkits afforded. While Wong et al. (2022) \citep{wong2022} mention the work practices that toolkits explicitly envision, we suggest that the toolkits' design decisions relating to content presentation and interaction modalities can also impact the work practices they support. For example, participants felt that the MIT Blindspots toolkit's provision of non-sequential navigation, where they could jump in and out of different parts, supported more iterative work processes. Our findings also show that toolkits' design decisions can further exacerbate a ``decontextualized approach to ethics''~[p. 14]\citep{wong2022}. For example, in the Nokia AI Design toolkit, while participants were able to input their answers directly into the tool, they were expecting custom or interactive responses that addressed what they had written. They felt that the outputs of the tool were too generic and even misleading in that they relied too heavily on toolkit users' ability to describe the system accurately and their integrity to describe it honestly.

\smallskip
\noindent Overall, our findings support several strands of previous research and extend them to new aspects and paradigms, as well as offering an understanding of how VSD can impact RAI toolkit design and how VSD values and RAI values align.

\subsection{Practical Implications}
\label{subsec:practical_implications}
Looking specifically at the toolkits' ability to operationalise the core concepts of VSD (i.e., working with stakeholders and embedding values into their work), the toolkits' design features support these aspects in various ways. We synthesized six design recommendation for creators of RAI toolkits. These recommendations are summarised in Figure \ref{fig:design-recommendations} in the Appendix.

\smallskip
\noindent\textbf{Encouraging Collaboration with Stakeholders through Open-Ended Cuing.}
Despite the majority of RAI tools targeting technical practitioners \citep{kaur2020, balayn2023, zhang2020, piorkowski2021}, several existing tools can support the inclusion of and collaboration with external or non-technical stakeholders through their design decisions. Supporting collaboration versus solo work has been a main point of discussion across the study. The ways in which the Nokia AI Design toolkit supports collaboration is through the open-ended and general nature of its cards, affording broad and unbiased ideation and opening several avenues for discussions. This can both support collaboration across teams, but can also lower the barrier-to-entry for non-technical stakeholders more explicitly and practically \citep{wong2022} as the cards can help spark new ideas they have not thought of before without being extremely technical or specific and thus less intimidating. While this open-endedness had identifiable disadvantages, such as an increased cognitive load to apply the toolkit to specific scenarios or technologies, it was the main design feature that supported collaboration found in the study. 

\smallskip
\noindent\textbf{Increasing Empathy through Examples, Case Studies and Mentioning Stakeholders.}
Empathy is crucial for ethical decision-making in engineering contexts \citep{hess2017}. Two main design features supported an increase in empathy and the consideration of diverse stakeholders' perspectives: the MIT Blindspots toolkit's provision of examples and case studies, and the Nokia AI Design toolkit's mention of numerous stakeholders in its cards. In the former, this led to participants' mentioning an improved ability to empathise and reflect on users' and stakeholders' experiences as they did not have to spend extra mental effort translating the information to their specific scenario. In the latter, participants were able to brainstorm considerations and steps that need to take place to build responsible AI that take into account a wider range of stakeholders and perspectives. Empathy and collaboration go hand-in-hand as collaboration fosters empathy which then fosters a more user-centred mindset in practitioners \citep{howcroft2021}.

\smallskip
\noindent\textbf{Supporting Iteration through Generalisability and Navigation.}
Our study surfaced two design features that could support iterative development (as is an inherent part of and recommended by VSD adapted to build AI systems \citep{umbrello2021}). Firstly, participants mentioned that the Nokia AI Design toolkit's open-ended nature made it suitable for both early-stage ideation during planning stages, and late-stage testing and evaluation phases. The toolkit could be used several time throughout the design process and the fact that results can be reloaded into the tool and compared can support constant improvement and iteration. This finding has parallels in previous work on AI where interpretations could differ depending on the stage a practitioner was involved in \citep{mitchell2019}. Secondly, participants mentioned that the MIT Blindspots toolkit's non-linear navigation allows them to jump back and forth as many times as needed between different phases without having to sequentially go through the cards every time. While this style of navigation might mean that certain cards are overlooked if toolkit users deem them irrelevant, it can support iterating through phases more practically.

\smallskip
\noindent\textbf{Supporting Reflectivity \& Meaningful Outcomes through Responsiveness and Feedback.}
Previous work has advocated for practitioner reflections \citep{morley2021} and identified RAI toolkits that explicitly call for such reflections \citep{wong2022}. In this study, we found that toolkits' design decisions, such as providing response and adaptive feedback to users' responses, can affect their ability and willingness to reflect. The provision of customised and adaptive feedback would make the outputs of such toolkits more meaningful and would allow toolkit users to reflect on their responses and improve their practices. The current lack of responsive feedback could lead to misleading outcomes where toolkit users feel their work is sufficient but it does not actually respect the required values. This deficiency also means that the effectiveness of the tool relies on its use and the discretion of its users in reporting their work. In the context of AI systems, previous works have already been weary of leaving too much up to the discretion of practitioners \citep{whittlestone2019c}, and without providing actionable and customised feedback, RAI toolkits risk following suit.

\smallskip
\noindent\textbf{Supporting Actionability \& Shared Knowledge through Creating Accessible Outcomes.}
Participants valued the Nokia AI Design toolkit providing them with accessible outcomes that they could refer back to easily and use in their work moving forward. It is rare that RAI toolkits provide outcomes in this form, despite recent findings showing that practitioners use AI ethics resources and their outcomes in a number of actionable ways \citep{yildirim2023}. With decks of cards especially and other toolkits such as the MIT Blindspots toolkit, users would have to provide their answers or outcomes in a separate form (e.g., on paper, on a Miro board) and there would be extra work needed to summarise or make sense of these outcomes. Given technical practitioners' resistance or reluctance to engage in value-based and ethics-related work \citep{manders2009}, seeing it as a burden or additional load \citep{morley2021}, providing accessible and actionable outcomes might encourage them to engage more. Providing outcomes in an easy-to-read form that is understandable by a variety of toolkit users can also serve as shared knowledge that helps teams establish a shared mental model of the outcomes produced \citep{cannon1993} and meaningful discourse \citep{sekiguchi2021} instead of appealing to one type of practitioner over the other (e.g., code for developers or sticky notes for designers).

\smallskip
\noindent\textbf{Reducing Cognitive Load through Designing with Values in Mind.}
Finally, mapping the Nokia AI Design toolkit to VSD values shows that RAI toolkits are able to implicitly respect a number of VSD values without being explicitly designed with these values in mind. Participants appreciated not having to think of more human-centred values during their work, rather preferring that the toolkit they are using considers these values for them implicitly in the recommendations and solutions it provides and the ideas it sparks. Such an approach could therefore help overcome practitioners' reluctance to engage in such work \citep{sweller1998, manders2009} and having to take on responsibilities outside of their roles to bridge disciplinary gaps across stakeholders \citep{deng2023}. While other interventions such as training practitioners to consider these values and understand their implications are certainly needed, implicitly supporting these values until practitioners are capable or willing to explicitly do so themselves can be extremely helpful. From the findings of this study where participants' did not reach a consensus with experts on certain cards, and previous work \citep{jakesch2022}, it becomes clear that ensuring toolkit users are aware of value definitions is crucial to avoid misunderstandings.

\subsection{Limitations and Future Work}
This study has three limitations that call for future research efforts. Firstly, the limited size of our participant sample reduces the generalisability of our results. Future work would benefit from testing the generalisability of these findings on larger samples. It is also worth noting that given recent findings that different groups prioritise and perceive values differently \citep{jakesch2022}, replicating this study with a different cohort besides early-career researchers as they might perceive the toolkits differently and react in other ways.

Secondly, we also acknowledge the relative homogeneity in researchers' backgrounds and the study's research context and realise that results might differ across different disciplines and regions. It is worth testing whether these links and implications also apply within other socio-cultural contexts and specific domains.

Finally, we opted in to test two RAI toolkits for the study practicalities. However, other RAI toolkits might be applicable. As such, future work should include: \emph{i)} a wider study with a larger number of RAI toolkits, \emph{ii)} a quantification of the exact effects of the different links established and their influence on each other, \emph{iii)} an exploration of how the presentation form or medium used by toolkits impacts their effects on users and the outcomes produced, and \emph{iv)} an exploration of how other theoretical frameworks besides VSD are operationalised. It is important to note that while steps were taken to improve comparability between the two toolkits used in this study, it is challenging to directly compare toolkits with different content and delivery mediums.

\section{Conclusion}
\label{sec:conclusion}
The aim of this study was to explore (i) the extent with which Responsible AI toolkits advocate for Value-Sensitive Design values in their content and recommendations and (ii) the extent with which different design features of these toolkits affects their ability to support VSD by promoting stakeholder collaboration and toolkit user learning. Through a qualitative approach involving workshops with 17 AI researchers using RAI toolkits, we highlighted relationships between RAI toolkits and VSD values, and explored the design features influencing stakeholder collaboration and user learning in RAI toolkits. Key findings include the facilitation of collaboration through open-ended cuing, increased empathy via examples and case studies, support for iteration through generalisability and navigation, meaningful outcomes through responsiveness and feedback, actionability and shared knowledge through accessible outcomes, and reduced cognitive load by implicitly integrating values in toolkit recommendations. These insights contribute to understanding the operationalisation of theoretical frameworks like Value Sensitive Design in Responsible AI toolkits, addressing the need for practical and user-friendly tools in the design of Responsible AI.

\begin{acks}
This work is partially funded by the Leverhulme Trust through the Leverhulme Centre for the Future of Intelligence [Award Number: RC-2015-067]. This work was also supported by The Alan Turing Institute’s Enrichment Scheme.
\end{acks}

\balance
\bibliographystyle{ACM-Reference-Format}
\bibliography{main}

\clearpage
\appendix

\section{Survey Questions}
\subsection{Pre-Workshop Survey}
\begin{enumerate}
    \item Please enter your name.
    \item Please enter your age.
    \item Please select your gender.
    \item Please select your role.
    \item How accurate are the following statements?
        \begin{itemize}
            \item I am familiar with how at least one type of AI-based system works
            \item I am familiar with how to build at least one type of AI-based systems
        \end{itemize}
    \item Which of the following applies to you (you can select more than one)?
        \begin{itemize}
            \item I've created an AI-based system before
            \item I've learned about AI-based systems through formal education (i.e. school or university)
            \item I've self-learned about AI-based systems through taking a course, studying online, or other activities for at least 6 months
            \item It has been my job to create AI-based systems for at least 6 months
            \item None of the above.
            \item Other (please specify)
        \end{itemize}
        \item How many years of experience do you have with AI-based systems (knowledge-based experience or hands-on experience)?
        \item If you would like to add any comments or extra information about your selections above then please do so here:
\end{enumerate}

\subsection{Post-Workshop Open-Ended Survey}
\begin{enumerate}
    \item Please describe how you incorporated the toolkits in brainstorming activities/steps/considerations for building responsible AI. Did you use any specific strategies?
    \item When and how would you see yourself using the toolkits during your current workflows/tasks/studies?
    \item Did the toolkits help you overcome any specific challenges?
    \item Were there any challenges where the toolkits did not help, or any support you needed that they did not provide?
    \item Do you have suggestions or areas for improvement with the toolkits?
\end{enumerate}

\section{Toolkits Considered for Inclusion}
\subsection{List of Toolkits Considered}

The toolkits considered from inclusion were obtained from Wong et al. (2022) \citep{wong2022}'s previous work and an online repository of practical tools for building responsible AI \citep{data-toolkits}, in addition to applicable RAI toolkits created in 2023, and are shown in Table \ref{tab:toolkits-reviewed}.

\begin{table*}[t!]
\centering
\resizebox{\textwidth}{!}{%
\begin{tabular}{|l|l|l|}
\hline
\multicolumn{1}{|c|}{\textbf{Number}} & \multicolumn{1}{|c|}{\textbf{Toolkit}} & \multicolumn{1}{c|}{\textbf{Source}} \\ \hline
1 & Aequitas & \url{http://aequitas.dssg.io/} \\ \hline
2 & AI Assessment Tool & \url{https://altai.ai4belgium.be/nl} \\ \hline
3 & AI Ethics Cards & \url{https://www.ideo.com/post/ai-ethics-collaborative-activities-for-designers} \\ \hline
4 & AI Explainability 360 Open Source Toolkit & \url{http://aix360.mybluemix.net/} \\ \hline
5 & AI Fairness 360 & \url{https://aif360.mybluemix.net/} \\ \hline
6 & AI Maturity Tool & \url{https://ai.digimaturity.vtt.fi/?lang=en} \\ \hline
7 & AI Meets Design Toolkit & \url{https://www.aixdesign.co/toolkit} \\ \hline
8 & AI System Ethics Self-Assessment Tool & \url{https://www.smartdubai.ae/self-assessment} \\ \hline
9 & Algorithmic Accountability Policy Toolkit & \url{https://ainowinstitute.org/aap-toolkit.pdf} \\ \hline
10 & Algorithmic Equity Toolkit (AEKit) & \url{https://www.aclu-wa.org/AEKit} \\ \hline
11 & Artifical Intelligence Impact Assessment & \url{https://ecp.nl/wp-content/uploads/2018/11/Artificial-Intelligence-Impact-Assesment.pdf} \\ \hline
12 & Audit AI & \url{https://github.com/pymetrics/audit-ai} \\ \hline
13 & Building an Algorithm Tool & \url{https://www.cdt.info/ddtool/} \\ \hline
14 & Cards for Humanity & \url{https://cardsforhumanity.idean.com/} \\ \hline
15 & Community Jury & \url{https://docs.microsoft.com/en-us/azure/architecture/guide/responsible-innovation/community-jury/} \\ \hline
16 & Consequence Scanning Kit & \url{https://www.doteveryone.org.uk/project/consequence-scanning/} \\ \hline
17 & Create your Own Datawalk & \url{https://data-en-maatschappij.ai/uploads/VUB-Datawalk-gids-ENG-v3-digitaal-paginas.pdf} \\ \hline
18 & Data Cards Playbook & \url{https://pair-code.github.io/datacardsplaybook/} \\ \hline
19 & Data Collection Bias Assessment & \url{https://data-en-maatschappij.ai/en/tools/tool-data-collection-bias-assessment-form} \\ \hline
20 & Data Ethics Canvas & \url{https://theodi.org/article/data-ethics-canvas/} \\ \hline
21 & Data Ethics Decision Aid & \url{https://dataschool.nl/deda/?lang=en} \\ \hline
22 & Data Ethics Framework & \url{https://www.gov.uk/government/publications/data-ethics-framework/data-ethics-framework} \\ \hline
23 & Data Ethics Guide & \url{https://www.cigref.fr/wp/wp-content/uploads/2019/02/Cigref-Syntec-Digital-Ethics-Guide-for-Professionals-of-Digital-Age-2018-October-EN.pdf} \\ \hline
24 & Deon Ethics Checklist & \url{http://deon.drivendata.org/} \\ \hline
25 & Design Ethically Toolkit & \url{https://www.designethically.com/toolkit} \\ \hline
26 & The Digital Ethics Compass & \url{https://ddc.dk/tools/toolkit-the-digital-ethics-compass/\#compass} \\ \hline
27 & Digital Impact Toolkit & \url{https://digitalimpact.io/toolkit/} \\ \hline
28 & Digital Inclusion Card Game & \url{https://data-en-maatschappij.ai/index.php?p=actions/asset-count/count\&id=184990} \\ \hline
29 & Dynamics of AI Principles & \url{https://aiethicslab.com/big-picture/} \\ \hline
30 & Ethical Explorers Pack & \url{https://ethicalexplorer.org} \\ \hline
31 & Ethical OS Toolkit & \url{https://ethicalos.org/} \\ \hline
32 & Ethics \& Algorithms Toolkit & \url{https://ethicstoolkit.ai/} \\ \hline
33 & Ethics Framework van Machine Intelligence Garage & \url{https://futurescope.digicatapult.org.uk/wp-content/uploads/2023/04/DC\_AI\_Ethics\_Framework-2021.pdf} \\ \hline
34 & Ethics Inc.: A Design Game for Ethical AI & \url{https://www.ethicsinc-ontwerpspel.nl/ethisch-ontwerpspel-voor-ai/} \\ \hline
35 & Ethics Kit & \url{http://ethicskit.org/tools.html} \\ \hline
36 & Fairlearn & \url{https://fairlearn.github.io/} \\ \hline
37 & Guidance Ethics & \url{https://ecp.nl/wp-content/uploads/2019/11/060-001-Boek-Aanpak-begeleidingsethiek-240165-binnenwerk-digitaal.pdf} \\ \hline
38 & Harms Modeling & \url{https://docs.microsoft.com/en-us/azure/architecture/guide/responsible-innovation/harms-modeling/} \\ \hline
39 & HAX Workbook and Playbook & \url{https://www.microsoft.com/en-us/haxtoolkit/workbook/} \\ \hline
40 & Intelligence Augmentation Design Toolkit & \url{https://futurice.com/ia-design-kit} \\ \hline
41 & InterpretML & \url{https://github.com/interpretml/interpret} \\ \hline
42 & Judgment Call & \url{https://docs.microsoft.com/en-us/azure/architecture/guide/responsible-innovation/judgmentcall} \\ \hline
43 & Lime & \url{https://github.com/marcotcr/lime} \\ \hline
44 & LinkedIn Fairness Toolkit (LiFT) & \url{https://github.com/linkedin/LiFT} ,  \url{https://engineering.linkedin.com/blog/2020/lift-addressing-bias-in-large-scale-ai-applications} \\ \hline
45 & MIT AI Blindspots & \url{https://aiblindspot.media.mit.edu/} \\ \hline
46 & Model Cards & \url{https://modelcards.withgoogle.com/about} \\ \hline
47 & Nokia AI Design Toolkit & \url{https://bell-labs.com/rai-prompts/} \\ \hline
48 & NLP CheckList & \url{https://github.com/marcotcr/checklist} \\ \hline
49 & People+AI Guidebook & \url{https://pair.withgoogle.com/guidebook/} \\ \hline
50 & \begin{tabular}[c]{@{}l@{}}Principles for Accountable Algorithms en \\ Social Impact Statement for Algorithms\end{tabular} & \url{https://www.fatml.org/resources/principles-for-accountable-algorithms} \\ \hline
51 & Product Impact Tool & \url{https://productimpacttool.org/nl/portal/} \\ \hline
52 & RAI Toolkit & \url{https://rai.tradewindai.com/} \\ \hline
53 & Responsible AI in Consumer Enterprise & \url{https://static1.squarespace.com/static/5d387c126be524000116bbdbt/5d77e37092c6df3a5151c866/1568138185862/Ethics-of-artificial-intelligence.pdf} \\ \hline
54 & Responsible AI Diagnostic & \url{https://pwc.qualtrics.com/jfe/form/SV\_0UF8EgBJdAnV8fr} \\ \hline
55 & SageMaker Clarify & \url{https://sagemaker-examples.readthedocs.io/en/latest/sagemaker\_processing/fairness\_and\_explainability/fairness\_and\_explainability.html} \\ \hline
56 & SDoC for AI/AI Servcie FactSheets & \begin{tabular}[c]{@{}l@{}}Arnold et al. (2019), "FactSheets: Increasing trust in AI services through supplier's declarations of conformity," \\ in IBM Journal of Research and Development, vol. 63, no. 4/5\end{tabular} \\ \hline
57 & The Tarot Cards of Tech & \url{https://www.artefactgroup.com/wp-content/uploads/2018/10/Artefact-Tarot-Cards-of-Tech\_downloadable.pdf} \\ \hline
58 & TensorFlow Fairness Indicators & \url{https://github.com/tensorflow/fairness-indicators} \\ \hline
59 & Unbias Toolkit & \url{https://unbias.wp.horizon.ac.uk/fairness-toolkit/} \\ \hline
60 & Weights and Biases & \url{https://wandb.ai/site} \\ \hline
61 & What If Tool & \url{https://pair-code.github.io/what-if-tool/} \\ \hline
\end{tabular}%
}
% \caption{The list of toolkits that were reviewer considered for use in this study and the source they originated from.}
\caption{The list of toolkits that were reviewed as part of the study, along with their source (at the time of writing)}
\label{tab:toolkits-reviewed}
\Description{A table of the toolkits considered for use in the workshops. Each toolkit is numbered and its name and source are included.}
\end{table*}

\subsection{Toolkit Exclusion Process}

\begin{figure*}[t!]
  \centering
  \includegraphics[width=\linewidth]{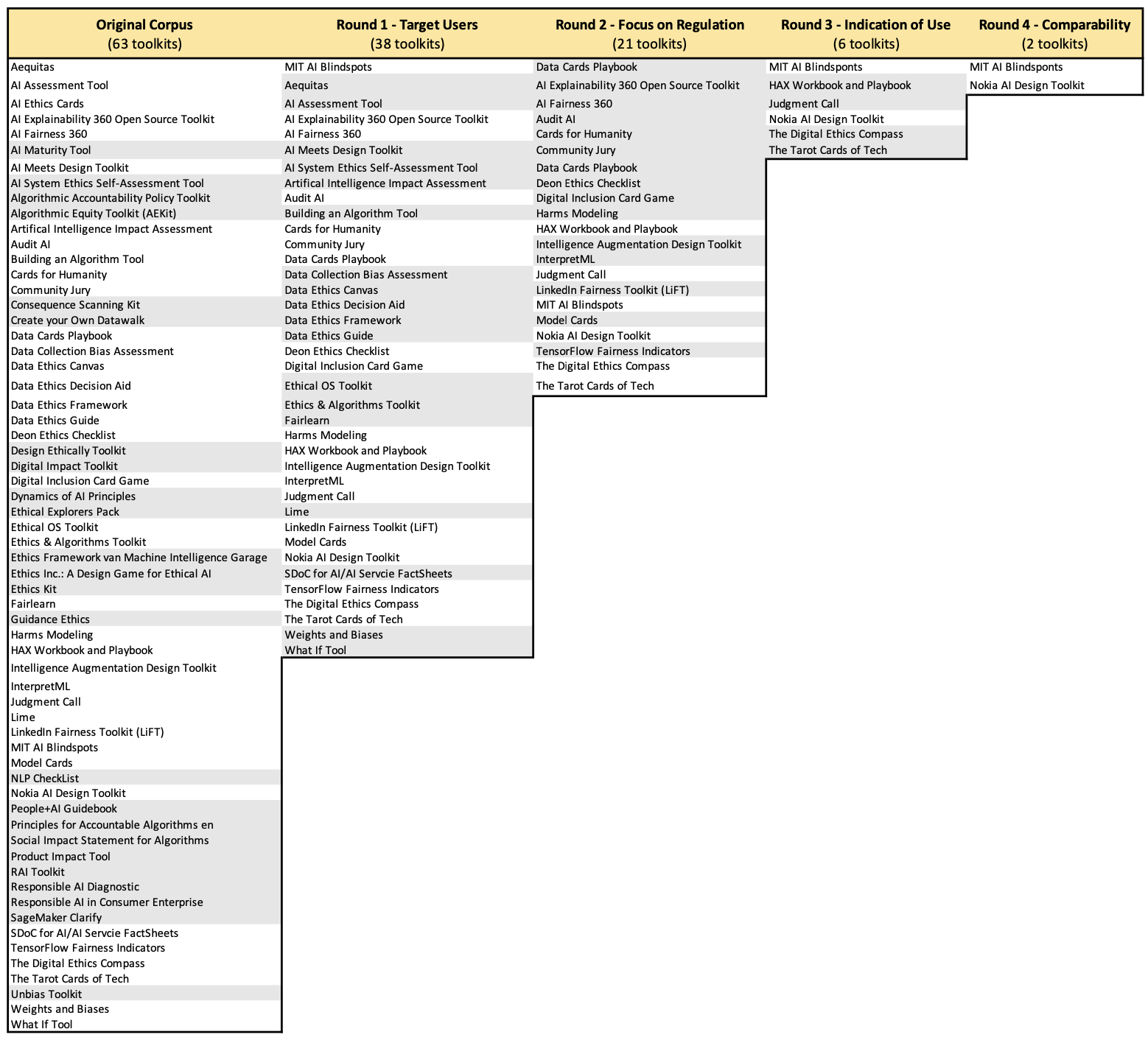}
  \caption{Lists of the toolkits considered during each round of the exclusion process as described in Section \ref{sec:methodology}. Each column shows the toolkits considered during that round. Toolkits highlighted in grey are the ones that were excluded in each round.}
  \label{fig:toolkits-considered}
  \Description{Lists of the toolkits considered during each round in the exclusion process. Each column shows the toolkits considered during that round as well as their total number and the exclusion criteria that was used during that round. Toolkits highlighted in grey are the ones that were excluded in each round.}
\end{figure*}

Figure \ref{fig:toolkits-considered} shows a list of toolkits considered during each round of the exclusion process, along with the total number of toolkits considered during each round and which toolkits were excluded.

\section{Design Recommendations}

\begin{figure*}[t!]
  \centering
  \includegraphics[width=0.8\linewidth]{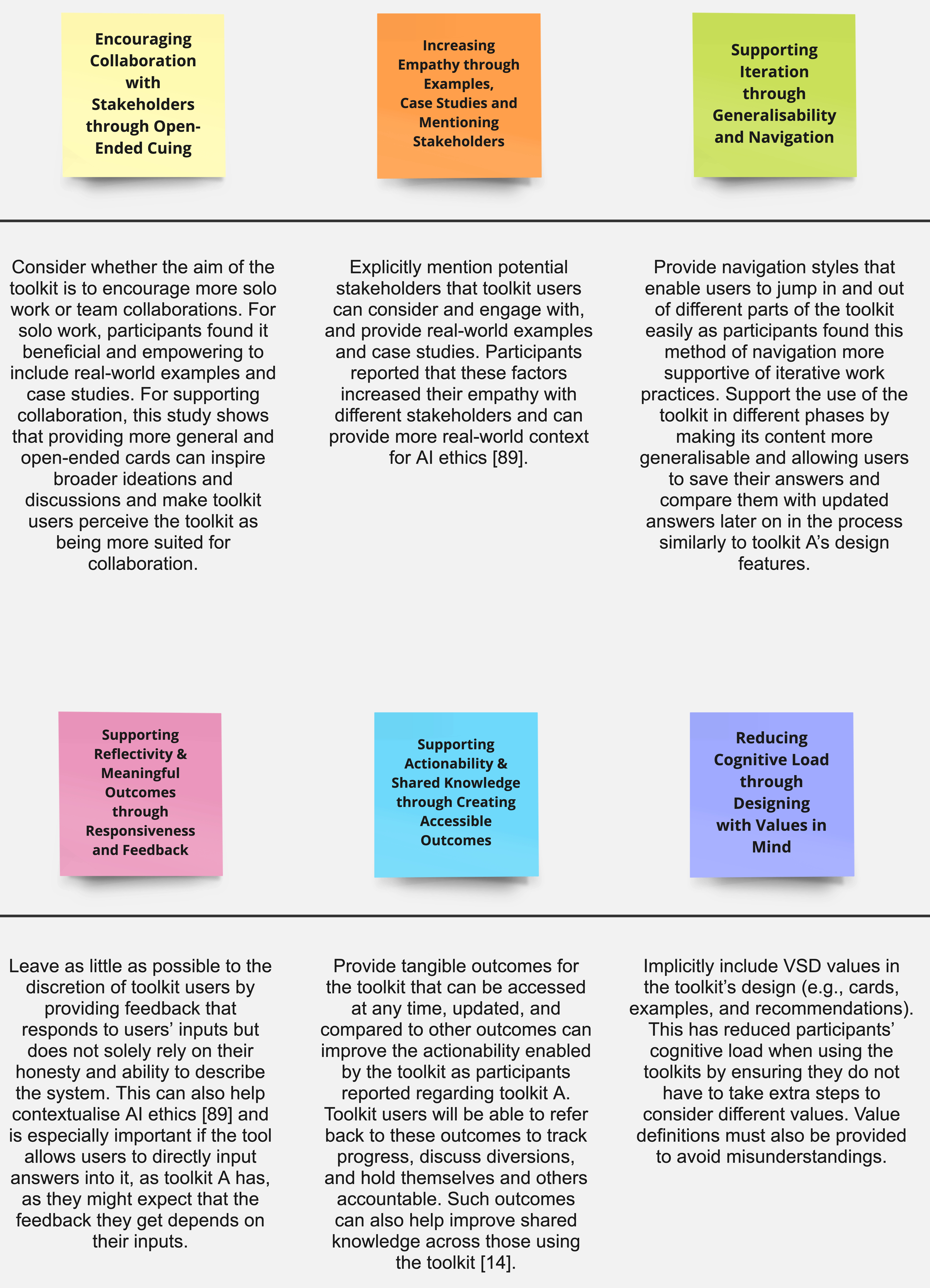}
  \caption{Six design recommendations we synthesised from our research. Each design recommendation is included in a sticky note with a detailed description included in the text under each sticky note.}
  \label{fig:design-recommendations}
  \Description{Six sticky notes showing the six design recommendations synthesised from the research. Underneath each sticky note is text describing each corresponding recommendation in detail.}
\end{figure*}

Figure \ref{fig:design-recommendations} summarises the six design recommendations for creators of RAI toolkits.

\end{document}